\documentclass{aa} 
\usepackage{graphicx} 
\usepackage{txfonts} 
\usepackage{colortbl} 
\definecolor{lightgray}{gray}{0.9}
\definecolor{heavygray}{gray}{0.5}
\usepackage[colorlinks=True, allcolors=blue]{hyperref} 
\usepackage{lscape}
\begin{document} 

\title{Stellar population astrophysics (SPA) with the TNG\thanks{Based on observations made with the Italian Telescopio Nazionale Galileo (TNG) 
operated on the island of La Palma by the Fundación Galileo Galilei of the INAF (Istituto Nazionale di Astrofisica) at the Spanish Observatorio 
del Roque de los Muchachos of the Instituto de Astrofisica de Canarias.
This study is part of the 
Large Program titled {\it SPA - Stellar Population Astrophysics:  the detailed, age-resolved
chemistry of the Milky Way disk} (PI: L. Origlia), granted observing time with HARPS-N and GIANO-B echelle spectrographs
at the TNG.}}
   \subtitle{The Arcturus Lab}

   \author{C. Fanelli \inst{1,2}
          \and
           L. Origlia\inst{2} 
          \and
          E. Oliva\inst{3} 
          \and
          A. Mucciarelli\inst{1,2} 
          \and
          N. Sanna\inst{3} 
          \and
          E. Dalessandro\inst{2}
          \and
          D. Romano\inst{2} 
          }

   \institute{Dipartimento di Fisica e Astronomia, Università degli Studi di Bologna, via Piero Gobetti 93/2, 40129, Bologna, Italy,
        \email{cristiano.fanelli3@unibo.it}
   \and
    INAF-Osservatorio di Astrofisica e Scienza dello Spazio, via Piero Gobetti 93/3, 40129, Bologna, Italy
    \and 
    INAF–Osservatorio Astrofisico di Arcetri, Largo Enrico Fermi 5, 50125 Firenze, Italy\\
             }
   \date{}
\abstract
{High-resolution spectroscopy in the near-infrared (NIR) is a powerful tool for characterising the physical and chemical properties of cool-star atmospheres.
The current generation of NIR echelle spectrographs enables the sampling of many spectral features over the full 0.9-2.4 $\mu$m range for a detailed chemical tagging.}
{Within the Stellar Population Astrophysics Large Program at the TNG, we used a high-resolution (R=50000) NIR spectrum of Arcturus acquired with the GIANO-B echelle spectrograph as a laboratory to define and calibrate an optimal line list and new diagnostic tools to derive accurate stellar parameters and chemical abundances.}
{We inspected several hundred NIR atomic and molecular lines to derive abundances of 26 different chemical species, including CNO, iron-group, alpha, Z-odd,  and neutron-capture elements. We then performed a similar analysis in the optical using Arcturus VLT-UVES spectra.}
{Through the combined NIR and optical analysis we defined a new thermometer and a new gravitometer for giant stars, based on the comparison of carbon (for the thermometer) and oxygen (for the gravitometer) abundances, as derived from atomic and molecular lines. 
We then  derived self-consistent stellar parameters and chemical abundances of Arcturus over the full $4800-24500 \, \AA$ spectral range and compared them with previous studies in the literature. 
We finally discuss a number of problematic lines that may be affected by deviations from thermal equilibrium and/or chromospheric activity, as traced by the observed variability of He~I  at $10830 \,\AA$. }
{}
\maketitle
\section{Introduction}
\label{intro}

The enhanced sensitivity of IR observations to intrinsically red (i.e. cool) and/or reddened (by dust extinction) objects make near-IR (NIR) spectrographs the ideal instrument for studying the physics, chemistry, and kinematics of cool giant and supergiant stars in galaxy fields as well as in star clusters. 

Cool giant and supergiant stars are among the brightest populations in any stellar systems and are easily observable at IR wavelengths out to large distances. They are also easy to detect in heavily reddened environments, such as the inner disk and bulge regions, where observations in the visual range are prohibitive.
These stars are important tracers of the star formation and chemical enrichment history of their hosts. 

High-resolution spectroscopy of these stars is crucial to obtain an exhaustive description of their detailed chemistry and nucleosynthesis.
Different chemical elements are synthesised in stars with different initial masses and thus released into the interstellar medium with different time delays with respect to the onset of star formation. The detailed chemical tagging of key elements is therefore crucial to constrain formation and chemical enrichment scenarios of the Milky Way and other nearby stellar system, in which these stars can be individually resolved.

In the past two decades high-resolution NIR spectroscopy has experienced a burst of activity in terms of newly commissioned spectrographs and stellar surveys. However, the precise identification and characterisation of the optimal atomic and molecular lines for abundance analysis, as well as their modelling 
over the entire NIR range, is still work in progress.

To this purpose, high-resolution spectroscopy in  both visual and NIR spectral ranges of suitable chemical calibrators is mandatory. 
Arcturus is such a calibrator for giant stars, and we present a comprehensive study in the YJHK NIR bands using the echelle spectrum at R=50000 that has recently been obtained with the GIANO-B spectrograph \citep{oli12a,oli12b,giano14,tozzi16}
at the Telescopio Nazionale Galileo (TNG). 

\begin{table*}[ht]
\centering
\caption{\label{reftab} 
 Stellar parameters and metallicity of Arcturus inferred from different optical and NIR studies.}
\begin{tabular}{lccccccc}
\hline
$T_{eff}$ & $log (g)$ & $\xi$ & $[Fe/H]$ & $Range$ & $Res$ & $References$\\
$K$ & dex & $km s^{-1}$ & dex & $Å$ & $\lambda/\Delta\lambda$ & \\
\hline
\rowcolor{lightgray}
$4283 \pm 39$  &  $1.55$  &  $1.61 \pm 0.03$ & $-0.55 \pm 0.07$ & $5370-7880$ & $45000^{*}$ & \cite{Fulbright06}\\   
$ 4290$  &  $1.55$  &  $1.70$ & $-0.50$ & $5000-7000$ & $50000^{**}$ & \cite{Ryde09}\\   
\rowcolor{lightgray}
$4286 \pm 30$ &  $1.66 \pm 0.05$  &  $1.74$ & $-0.52 \pm 0.02$ & $5000-9300$ & $100000^{**}$ & \cite{Ramirez-Prieto11} \\   
$4275 \pm 50$  &  $1.70 \pm 0.10$  &  $1.85 \pm 0.05$ & $-0.52 \pm 0.04$ & $15100-17000$ & $22300^{**}$ & \cite{Smith13}\\   
\rowcolor{lightgray}
$4286 \pm 50$  &  $1.66 \pm 0.10$  &  $1.70 \pm 0.05$ & $-0.57 \pm 0.04$ & $15100-17000$ & $22300^{**}$ & \cite{Shetrone15}\\   
$4286 \pm 35$  &  $1.64 \pm  0.06$  &  $1.22 \pm 0.12$ & $-0.69 \pm 0.06$ & $9300-13100$ & $28000^{***}$ & \cite{Kondo19}$^{a}$\\   
\rowcolor{lightgray}
$4286 \pm 35$  &  $1.64 \pm  0.06$  &  $1.20 \pm 0.11$ & $-0.49 \pm 0.04$ & $9300-13100$ & $28000^{***}$ & \cite{Kondo19}$^{b}$\\   

\hline\hline
\end{tabular}
   \tablefoot{\\
   \tablefoottext{a}{Using VALD3 line list: http://vald.astro.uu.se}\\
   \tablefoottext{b}{Using \cite{MB99} line list}\\
   \tablefoottext{*}{Spectrum observed with the $0.6$ m CAT telescope at the Lick Observatory and the Hamilton spectrograph}\\
   \tablefoottext{**}{Spectrum from \citet{Hinkle05}}\\
   \tablefoottext{***}{Spectrum observed with WINERED mounted at the $1.3$ m Araki Telescope at Koyama Astronomical Observatory.}}
\end{table*}

Arcturus ($\alpha \, Boo$, $HR5340$, $HIP69673$, or $HD124897$) is a luminous, nearby K1.5 IIIp giant star that is often used as a calibrator in chemical studies of cool stellar populations. 
Differential chemical analysis of giant stars relative to Arcturus can indeed largely minimise most of systematic errors due to atmospheric parameters \citep[e.g.][]{McWilliam&Rich94,Worley09,Alves-Brito2010,Ramirez-Prieto11}.

However, it is challenging to take a spectrum of Arcturus because of its apparent ultra-bright luminosity. 
Most of the chemical studies of Arcturus are based on  high-resolution spectroscopy.  
The reference Arcturus spectrum covering the entire spectral range from the UV to the IR is the one made available by \citet[][and references therein]{Hinkle05}. 
This spectrum has been built using three different instruments: the Space Telescope Imaging Spectrograph (STIS) mounted on board of the Hubble Space Telescope in the  $1000$-$3000$ Å range, the echelle optical spectrograph  in the  $3100$-$9000$ Å range, and the Fourier transform spectrometer in the $0.9$-$5$ $\mu m$ range mounted at the Kitt Peak National Observatory (KPNO) $4$ m telescope. 

\citet{Ryde09} used a portion of the \citet{Hinkle05} H-band spectrum, from $15326$ to $15705$ Å, to study several clean molecular lines of CO, CN, and OH and derive C, N, and O abundances.
\citet{Ramirez-Prieto11} provided atmospheric parameters and abundances for several metals by mostly using the \citet{Hinkle05} optical spectrum and the line list by \citet{Asplund09}. 
\cite{Smith13} and \cite{Shetrone15} used the H-band portion  of the \citet{Hinkle05} spectrum and their detailed line list prepared for the Sloan Digital Sky Survey III Apache Point Galactic Evolution Experiment (APOGEE) 
\citep{APOGEE} to provide stellar parameters, Fe, CNO and other elemental abundances.

Arcturus was also studied by  \citet{Fulbright06} using an optical spectrum taken with the Hamilton spectrograph at the $0.6$m CAT telescope of the Lick Observatory. \citet{Kondo19} have analysed a ZYJ spectrum of Arcturus at R$\simeq 28000$ taken with the WINERED spectrograph at the $1.3$ m Araki Telescope at Koyama Astronomical Observatory. They derived Fe~I abundances and microturbolence using two different line lists: the Vienna Atomic Line Database (VALD3) \citep{Ryabchikova15}, and the public line list provided by \citet{MB99}. 

Finally, we mention the works by \citet{Maas17,Dorazi11} and \citet{Overbeek16}, who discussed the Arcturus  abundances of P, Y, and Dy, respectively. Table~\ref{reftab} lists the stellar parameters and metallicity [Fe/H] of Arcturus inferred from different optical and NIR studies.

This paper is organised as follows. In Section 2 we describe the observation and the data reduction of the Arcturus spectrum. In Section 3  we discuss the method we adopted for spectral analysis, and in Section 4 we describe the procedure we used to determine the stellar parameters for Arcturus using new NIR diagnostics. In Section 5 we report the results of our chemical analysis in the optical and NIR range, and in Section 6 we compare them with those from previous studies and draw our conclusions.

\section{Observations and data reduction} 

Arcturus was observed on July $2^{}$, 2018, with GIANO-B, the high-resolution (R=$50000$) NIR ($9500$-$24500$ Å) spectrometer at the TNG \citep{Origlia14}. The observation was part of the Large Program called \textit{\textup{Stellar population astrophysics: detailed age-resolved chemistry of the Milky Way disk}} (PI: L. Origlia). 
Spectra were collected using a grey filter that attenuates the  light by about 5 magnitudes. Nodding was used to optimise the subtraction of the background and other detector patterns: we collected several pairs of exposures with the star alternatively positioned at 1/4 (position A) and 3/4 (position B) of the slit length.
The integration time was $60$ seconds per A,B position, with a mean seeing of $\sim 0.7''$. 

The raw spectra were reduced using the data reduction pipeline software GOFIO \citep{gofio}, which processes calibration (darks, flats, and U-Ne lamps taken in daytime) and scientific frames. The main feature of the GOFIO data reduction is the optimal spectral extraction and wavelength calibration based on a physical model of the spectrometer that accurately matches instrumental effects such as variable slit tilt and order curvature over the echellogram \citep{Oli18}. The data reduction package also includes bad pixel and cosmic removal, sky and dark subtraction, flat-field and blaze correction.

The spectrum was corrected for telluric absorption using the spectra of an O-type standard star taken at different air masses during the same night. The normalised spectra of the telluric standard taken at low and high air-mass values were combined with different weights to match the depth of the telluric lines in the Arcturus spectrum.
Figs. \ref{yband}, \ref{jband}, \ref{hband}, and \ref{kband} in the appendix show the   rest-frame normalised spectra corrected for telluric absorption.
The average signal-to-noise ratio of the reduced and telluric-corrected spectrum 
is about 150 per pixel.

We also analysed two optical spectra of Arcturus retrieved from the ESO archive, in order to cross-check chemical abundances over the widest possible spectral range. These optical spectra were collected with the high-resolution spectrograph UVES at the ESO Very Large Telescope (VLT) at a resolution of R$\sim80000$, using the CD3 Red Arm 580 and CD4 Red Arm 860, which cover the $4800$ - $6800$ $\AA$ and $6700$ - $10400$ $\AA$ wavelength ranges, respectively.

\section{Spectral analysis}\label{sec:Spec_analysis}

Accurate and precise stellar parameters and chemical abundances of Arcturus were determined by means of spectral synthesis technique applied to the observed spectra. Synthetic spectra were computed by using the radiative transfer code {\tt TURBOSPECTRUM} \citep{Alvarez&Plez98,Turbospec} with MARCS models atmospheres \citep{MARCS1}, 
the atomic data from VALD3 and the most updated molecular data from the website of B. Plez,  \url{https://www.lupm.in2p3.fr/users/plez/}. 
The synthetic spectra were convoluted with a Gaussian function in order to reproduce the observed broader profile that corresponds to an equivalent resolution of $32,000$. The additional broadening is mainly due to the macro-turbulence velocity ($\approx 6 \, km s^{-1}$) because the projected rotational velocity of Arcturus is negligible ($\xi_{rot}=2.4 \, km s^{-1}$, \citealt{Gray81}).

For the abundance analysis, we used a selected list of C~I, Na~I, Mg~I, Al~I, Si~I, P~I, S~I, K~I, Ca~I, Sc~I, Ti~I, V~I, Cr~I, Mn~I, Fe~I, Fe~II, Co~I, Ni~I, Cu~I, Zn~I, Y~I; Y~II, Ce~II, Nd~II, and Dy~II atomic lines and CO, OH, CN, and HF molecular lines. Each line was carefully checked against possible blending with close contaminants. For this purpose, we developed a code called {\tt TurboSLine} that  identifies as potential contaminants any atomic or molecular j-th line whose centroid $\lambda_j$ is within one full width at half maximum  (FWHM) from the centroid  $\lambda_i$ of the analyzed i-th line.\\
For each of these potential contaminants, we computed the theoretical line equivalent width (EW) and the amount of contamination using the following approximation:
\begin{equation}
    C_j = EW_j \times \bigg(1-\frac{|\lambda_i - \lambda_j |}{FWHM_j}\bigg)
.\end{equation}
If $\sum C_j > 0.1\times EW_i$, the i-th line was classified as blended, and it was not normally used for abundance analysis.\\

Some other lines were later on rejected because they are contaminated by the wings of nearby strong photospheric and/or by deep telluric lines by visual inspection.
In our chemical analysis we also rejected strong lines because of the uncertainty in the modelling of their wings, non-local thermal equilibrium (NLTE), and chromospheric effects, etc.

Tables A.1-A.4\footnote{Tables A.1-A.4 are only available in electronic form at the CDS via anonymous ftp to \url{cdsarc.u-strasbg.fr} (130.79.128.5) or via \url{http://cdsweb.u-strasbg.fr/cgi-bin/qcat?J/A+A/}} provide the complete list of optical and NIR atomic and molecular lines used for the abundance analysis. 
For the computation of the chemical abundances we used {\tt SALVADOR}, a tool developed by A. Mucciarelli \citetext{priv.\ comm.}, which  performs a $\chi ^2$ minimisation between observed and synthetic spectra while the normalisation of the observed spectrum around each line is optimised interactively. 

As a further check, we also computed the line EWs and derived the corresponding abundances. The latter were found to be practically coincident with those obtained from the spectral synthesis, with eventually only a slightly higher dispersion, likely because lines with some impurity provide slightly more uncertain abundances when the EW method is used.

\section{Stellar parameters}\label{param}

\begin{figure}[!h]
    \centering
    \includegraphics[scale=0.35]{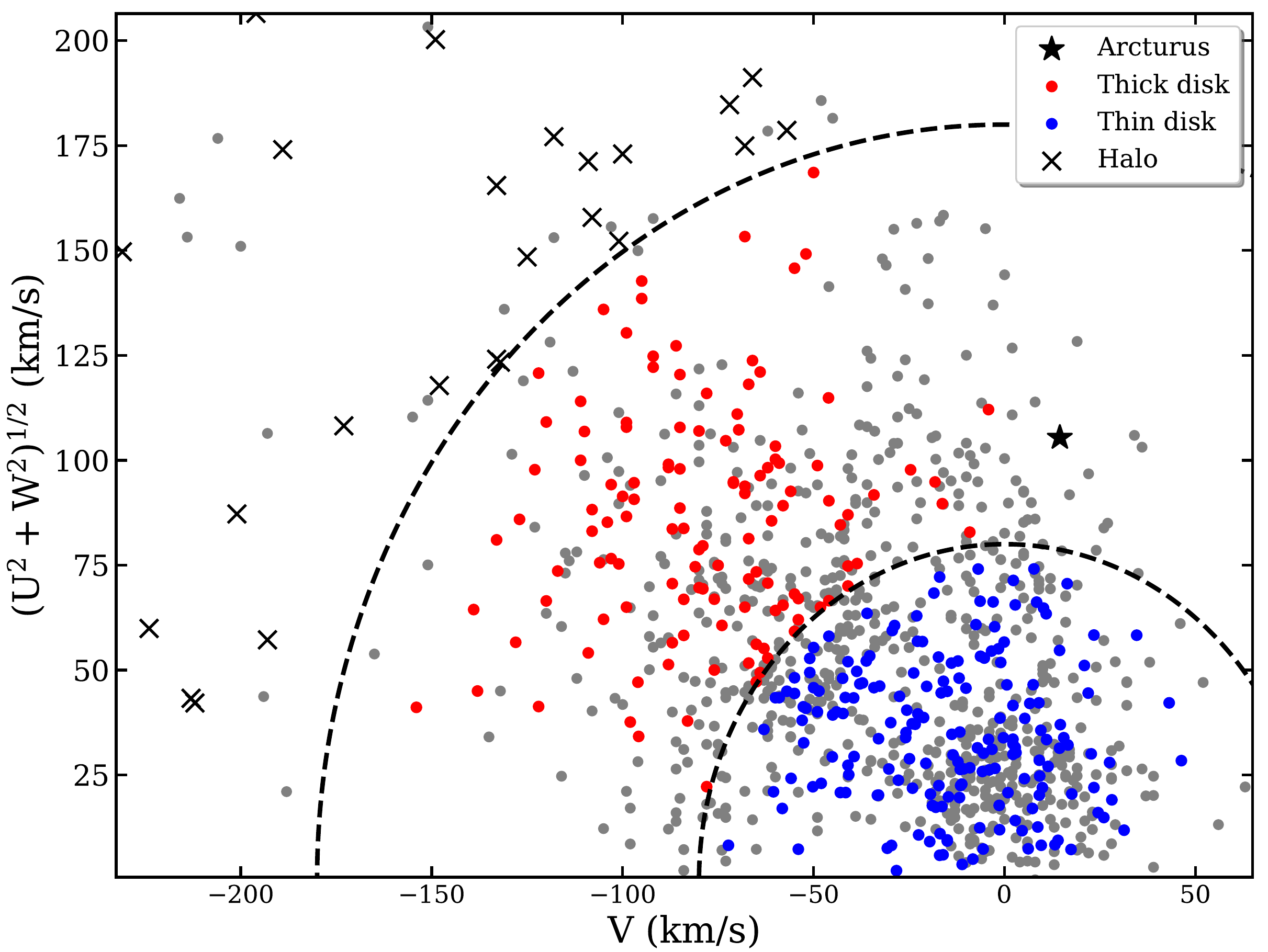}
    \caption{Toomre diagram for thin-disk (blue dots), thick-disk (red dots), and halo (black crosses) stars from \citealt{Reddy03} and \citealt{Reddy06} and for the disk stars of \citealt{bensby14} (grey dots).
    The black star marks the position of Arcturus, and the dashed lines delineate constant total space velocities with respect to the LSR of V$_{tot}=85$ and $180$ km $s^{-1}$, respectively \citep[e.g.][and references therein]{nissen04}.}
    \label{toomre1}
\end{figure}

Arcturus belongs to a kinematic group of several dozen old stars \citep{Eggen71}. 
Its metallicity ([Fe/H]$\simeq -0.5$~dex) and some enhancement of the alpha elements suggest that it likely formed in the thick disk of our Galaxy \citep{Ramirez-Prieto11,bensby14}, although an extragalactic origin has also been proposed \citep{Navarro04}. 
From a kinematic perspective, the Arcturus total velocity $v_t=\sqrt(U^2+V^2+W^2)=106$~km~s$^{-1}$ and its location in the Toomre diagram (see Fig.~\ref{toomre1}) suggest that it is a thick disk star \citep{bensby14}. 

Previous determinations of the Arcturus stellar parameters (cf. Table~\ref{reftab}) suggested temperatures in the 4275-4290~K range, log(g) in the 1.55-1.70 dex range, and microturbulence in the 1.20-1.85 $km s^{-1}$ range. Using the Dartmouth web-tool at {\it http://stellar.dartmouth.edu} \citep{Dotter08}, we computed isochrones with [Fe/H]=$-0.5$, [$\alpha$/Fe]=$+0.20$ and different ages.
At the bolometric luminosity of Arcturus ($L_{bol}=174\pm7~L_{\odot}$, \citet{Smith13}, see also \citet{Ramirez-Prieto11}), we found that old ($\ge$10 Gyr) ages are consistent with an effective temperature T$_{eff}$ in the 4260-4310 K range and gravity log(g) in the 1.60-1.70 dex range.

\begin{figure}[h]
    \centering
    \includegraphics[scale=0.33]{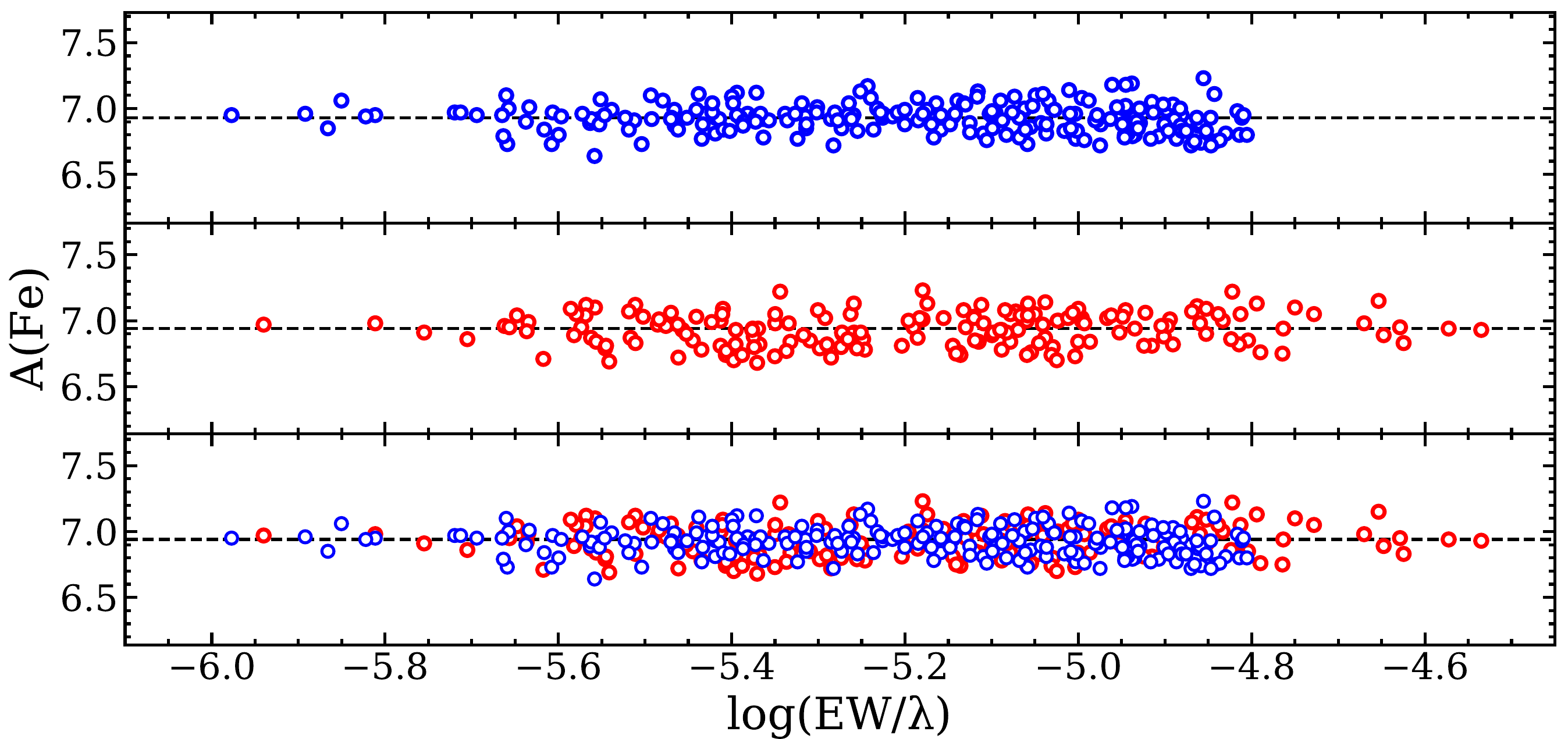}
    \caption{Iron abundances as a function of the reduced EW for all the measured optical (blue circles) lines in the UVES spectra and NIR (red circles) lines in the GIANO-B spectrum. The dotted line marks the best-fit median abundance.
    \label{Fe_micro}}
\end{figure}

We used these photometric ranges for T$_{eff}$ and log(g) to also constrain the microturbulence velocity range with the standard approach of minimising the slope between the iron abundance and the reduced EW of the measured lines $log(EW/\lambda)$ \citep[see the discussion in][]{Muccia_micro}. We find microturbulence velocities in the 1.50 and 1.70 $km s^{-1}$ range. 
We finally adopted the value of $\xi=1.60 \pm 0.05 \, kms^{-1}$ that best minimizes any trend between the abundances and the reduced EW of about 400 iron lines distributed over the $4800-23400 \, \AA$ spectral range, as shown in Fig.~\ref{Fe_micro}. This value of microturbolence also minimises the trend when the optical and the NIR lines are taken separately.

As described in the following, the photometric values of T$_{eff}$ and log(g) were fine-tuned using the spectroscopic data. 

\subsection{C-thermometer}\label{subsec:C-thermometer}

We defined a new powerful diagnostic tool to derive $T_{eff}$ in oxygen-rich cool stars, based on the balance between the carbon abundance, derived from atomic lines (see Table~\ref{selected_carbon_lines}), and molecular CO roto-vibration transitions. Hereafter, we refer to this method as the C-thermometer.\\

\begin{table}[!h]
\centering
\caption{\label{selected_carbon_lines} Measurable atomic carbon lines in the NIR spectrum of Arcturus.}
\begin{tabular}{lccc}
\hline
$\lambda_{air}$ & $\chi$ & log(gf)$^{(a)}$ & Note  \\
$ \AA $ & eV &  &  \\
\hline
\rowcolor{lightgray}
$8727.140$ & $1.26$ & $-8.165$ & Forbidden  \\
$9658.435$ & $7.49$ & $-0.280$ & NLTE  \\
\rowcolor{lightgray}
$10683.080$ & $7.48$ & $+0.079$ & NLTE  \\
$10685.340$ & $7.48$ & $-0.272$ & NLTE  \\
\rowcolor{lightgray}
$10691.245$ & $7.49$ & $+0.344$ & NLTE  \\
$10707.320$ & $7.48$ & $-0.411$ & NLTE  \\
\rowcolor{lightgray}
$10729.529$ & $7.49$ & $-0.420$ & NLTE  \\
$11748.220$ & $8.64$ & $+0.375$ & NLTE  \\
\rowcolor{lightgray}
$11753.320$ & $8.65$ & $+0.691$ & NLTE  \\
$11754.760$ & $8.64$ & $+0.542$ & NLTE  \\
\rowcolor{lightgray}
$17234.463$ & $9.70$ & $+0.293$ & LTE  \\
$17448.535$ & $9.00$ & $+0.012$ & LTE   \\
\rowcolor{lightgray}
$17672.039$ & $7.95$ & $-1.974$ & LTE  \\
$17768.910$ & $9.71$ & $+0.420$ & LTE  \\
\rowcolor{lightgray}
$17793.158$ & $9.71$ & $-0.045$ & LTE  \\
\hline\hline
\end{tabular}
\tablefoot{\\
   \tablefoottext{a}{For all the tabulated lines we used the NIST log(gf) values as listed in the VALD3 database.}\\
    }
\end{table}

\begin{figure}[!h]
    \centering
    \includegraphics[width=\hsize]{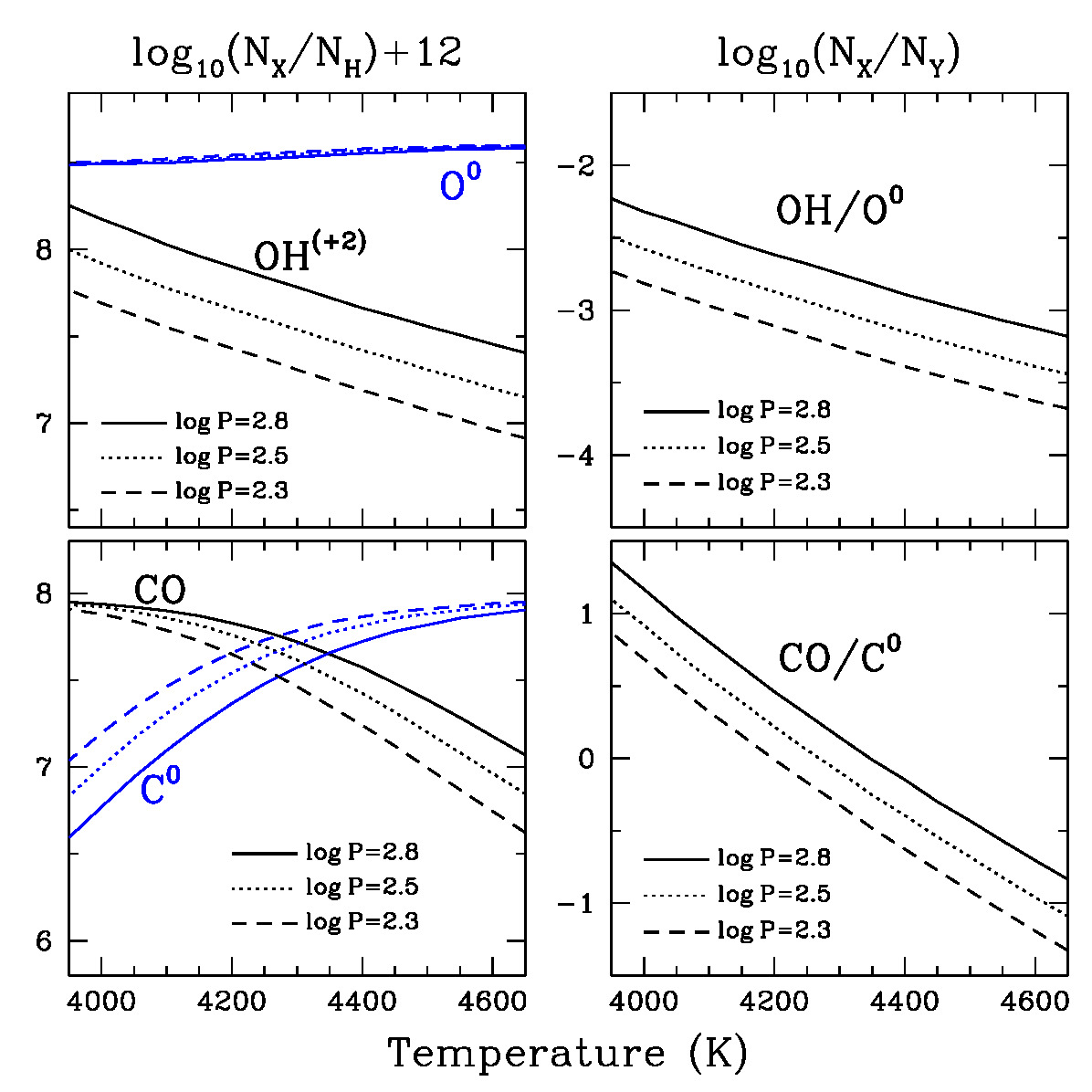}
    \caption{Selected pressure values are representative of the line-forming regions for surface gravities log(g)=2.0, 1.5, and 1.0
    \label{fig_saha_equilibrium}}
\end{figure}

The basic principle of this thermometer follows from the very high dissociation potential of the CO molecule (11.1 eV). Thus the CO/C$^0$ relative abundance of carbon in molecular (CO) and atomic (C$^0$) form has a very strong exponential dependence on temperature. Like for other diatomic molecules, the abundance ratio also depends linearly on the gas pressure (i.e. on gravity); but this effect is far weaker. Figure~\ref{fig_saha_equilibrium} shows the behaviour of the Saha equilibrium abundances for gas temperatures and pressures relevant for this work.\\
When the spectrum is modelled, CO and C~I lines must provide the same abundances. If this does not happen, it is because the model temperature is incorrect and must be tuned until the two abundances match.
However, the analysis is complicated by the fact that some C~I lines may show significant departure from LTE (see e.g. \citealt{Fabbian06,Takeda13}), while CO lines form under LTE conditions \citep[e.g.][]{Hinkle&Lambert75,Ryde09}.
For example, \citet{Takeda13} showed that the  $10683/10685/10691 \, \AA$ multiplet in the Y band requires NLTE correction for a star like Arcturus. In particular, the $10691 \, \AA$ line needs to be corrected by $\Delta A(C) = -0.23$ dex. \\
NLTE corrections are sensitive to the adopted stellar model and depend on temperature and gravity, hence they cannot be safely used to define a reliable thermometer. It is therefore desirable to use only those C~I lines that are not affected by NLTE effects.

\begin{figure}[!h]
    \centering
    \includegraphics[scale=0.37]{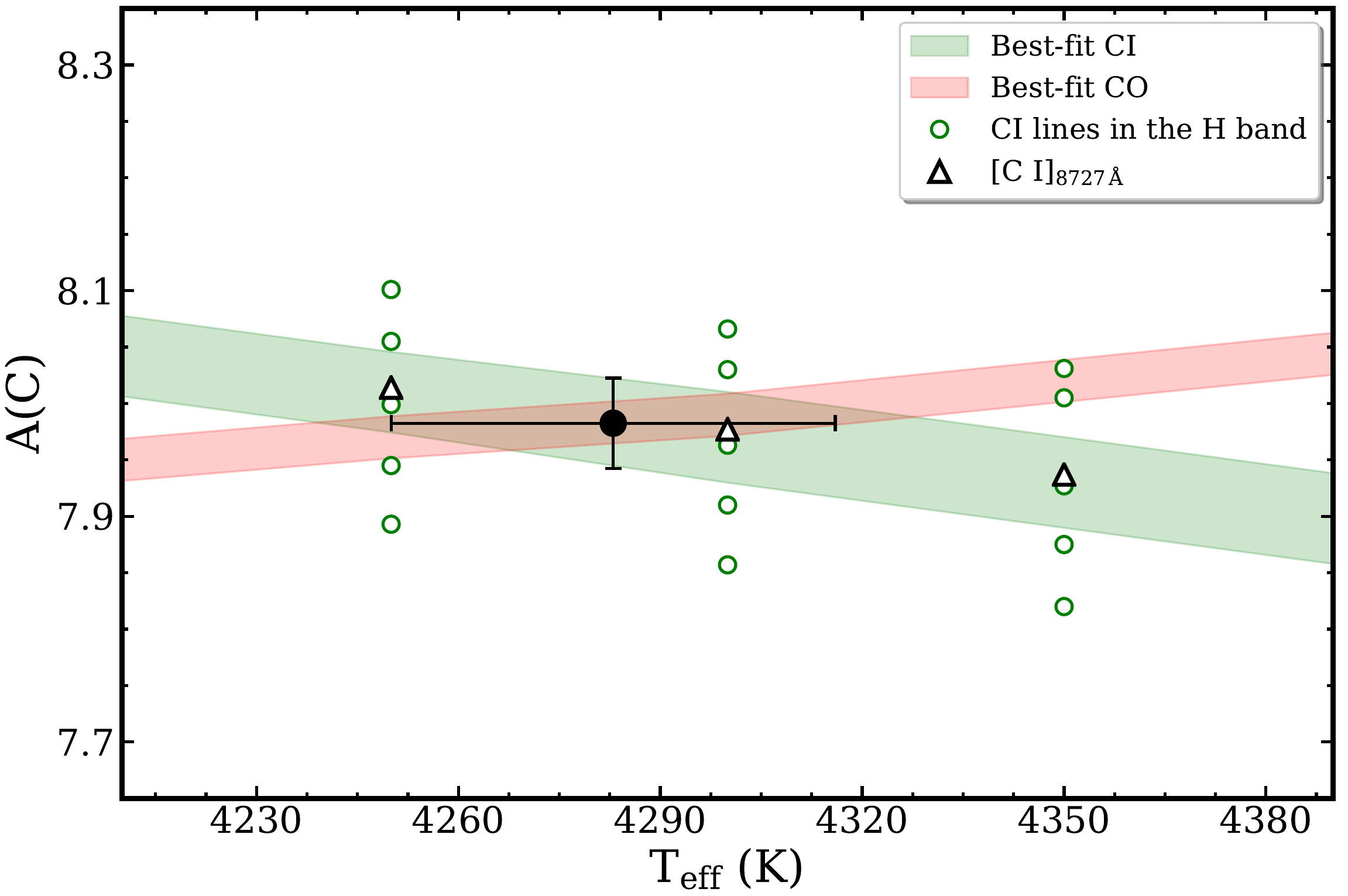}
    \caption{C abundances as a function of temperature from the C~I lines in the H band (green circles), and the forbidden [C I] line at $8727.14 \, \AA$ 
    (empty triangle). The shaded green region is the corresponding best-fit $\pm 1\sigma$ trend of the C~I lines, while the shaded red region is the best-fit $\pm 1\sigma$ trend  of the CO lines. The large black dot 
    marks the intersection of the two curves, and its x-coordinate 
    provides the best-fit temperature.
    \label{Ctermo3}}
\end{figure}

For this purpose, given that forbidden lines do not suffer from NLTE \citep[e.g.][]{alexeeva15}, we used the  [CI] at $8727.14 \, \AA$  measurable in our UVES red spectrum to derive a proxy of the atomic carbon abundance in LTE. Then, we computed LTE abundances for all the C~I lines measurable in our GIANO-B spectrum (Table~\ref{selected_carbon_lines}), and we checked their abundances against that from the [C I] line.
We found that the C~I lines in the H band with excitation potentials above 9 eV provide similar C abundances, and the C~I lines in the Y and J bands with lower excitation potentials give LTE abundances that are systematically ($\sim$ 0.3 dex) higher. The reason most likely are NLTE effects. We therefore used only the C~I lines in the H band for the C-thermometer.

Fig.~\ref{Ctermo3} shows the variation in carbon abundance from C~I and CO lines in the Arcturus GIANO-B spectrum as a function of $T_{eff}$. Both diagnostics are very sensitive to $T_{eff}$ , but have opposite trends. The two curves intersect at T$_{eff}=4283 \pm 33$~K. The quoted error of $\pm$33 ~K corresponds to a $\pm1\sigma$ variation in the derived C abundances from CI and CO.
The C-thermometer is virtually independent of the other parameters within the uncertainties. 
Variations of $\pm 0.05$ $km s^{-1}$ in microturbulence velocity or $\pm 0.06$ dex in log(g) have a negligible effect on temperature ($\le15$ K) and C abundance ($\le$0.02 dex). Even for a stronger variation of log(g) (up to 0.2 dex) and $\xi$ (up to 0.2 $kms^{-1}$) does the corresponding variation in temperature lie within the error.
Interestingly, the C-thermometer also works if the molecular carbon abundance is derived from spectral synthesis of the $\Delta v$=3 CO band-heads in the H band because it is fully consistent with the one derived from individual CO roto-vibration lines (see Sect.~\ref{cno}).

\begin{figure}[t]
    \centering
    \includegraphics[scale=0.33]{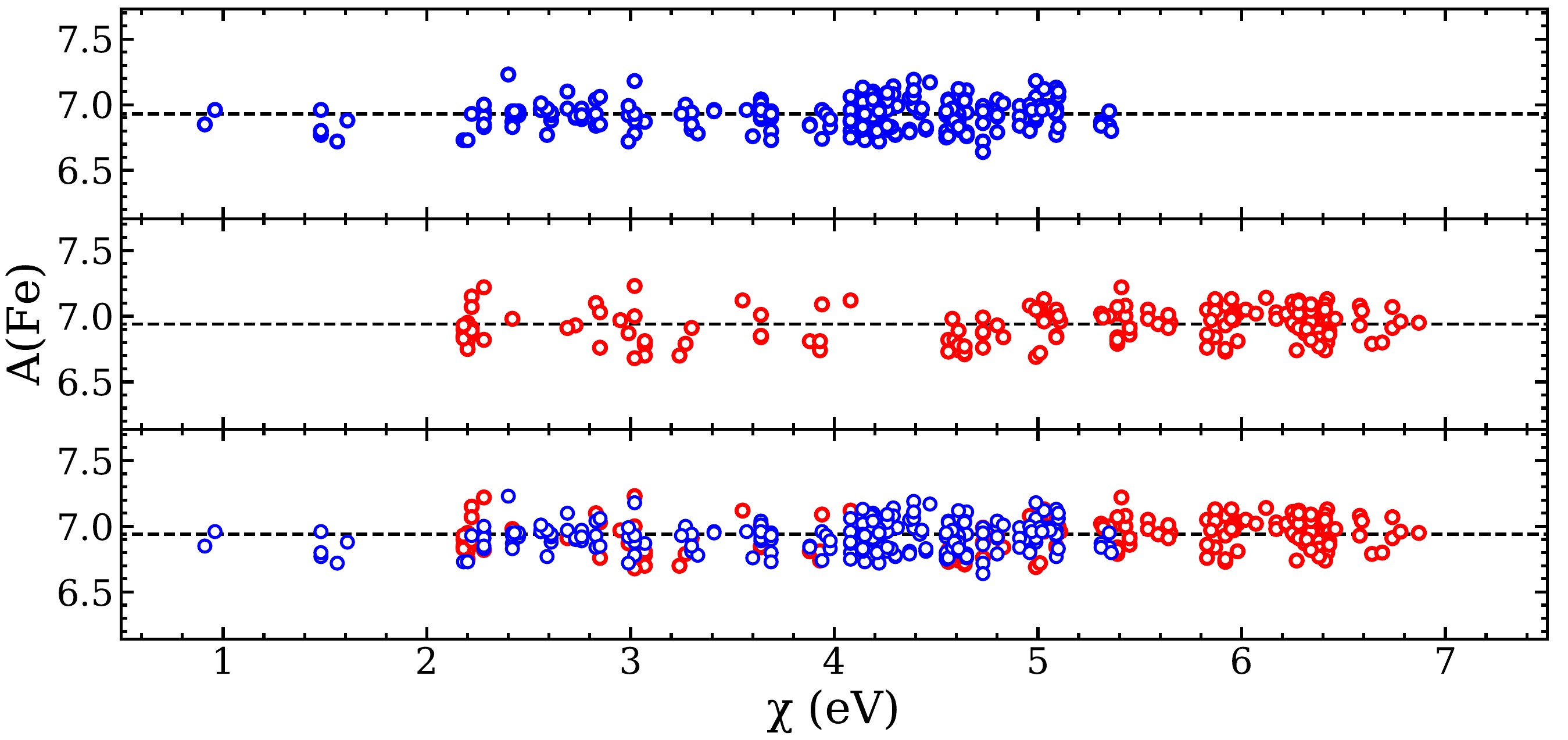}
    \caption{Iron abundances from neutral lines in the optical UVES spectra (blue circles) and in the NIR GIANO-B spectrum (red circles) as a function of their excitation potential. The dotted line marks the derived best-fit median iron abundance.
    \label{Fe_vs_chi}}
\end{figure}

Noticeably, the temperature of 4283 $\pm 33$ K derived from the C-thermometer 
also allows us to minimise any trend between iron abundances from neutral lines and their excitation potential within the errors, as shown in Fig.~\ref{Fe_vs_chi}, which is the standard spectroscopic method for inferring the effective temperature.

\subsection{O-gravitometer}\label{subsec:O-gravitometer}

The relative abundance of OH and atomic oxygen ($O^0$) depends linearly on the gas pressure (i.e. on gravity), while it has a weak dependence on temperature (see Figure~\ref{fig_saha_equilibrium}) because of the low dissociation potential of the OH molecule (4.4 eV). Therefore, the $OH/O^0$ ratio can be used to estimate the gravity when the temperature is constrained.

\begin{figure}[h]
    \centering
    \includegraphics[scale=0.395]{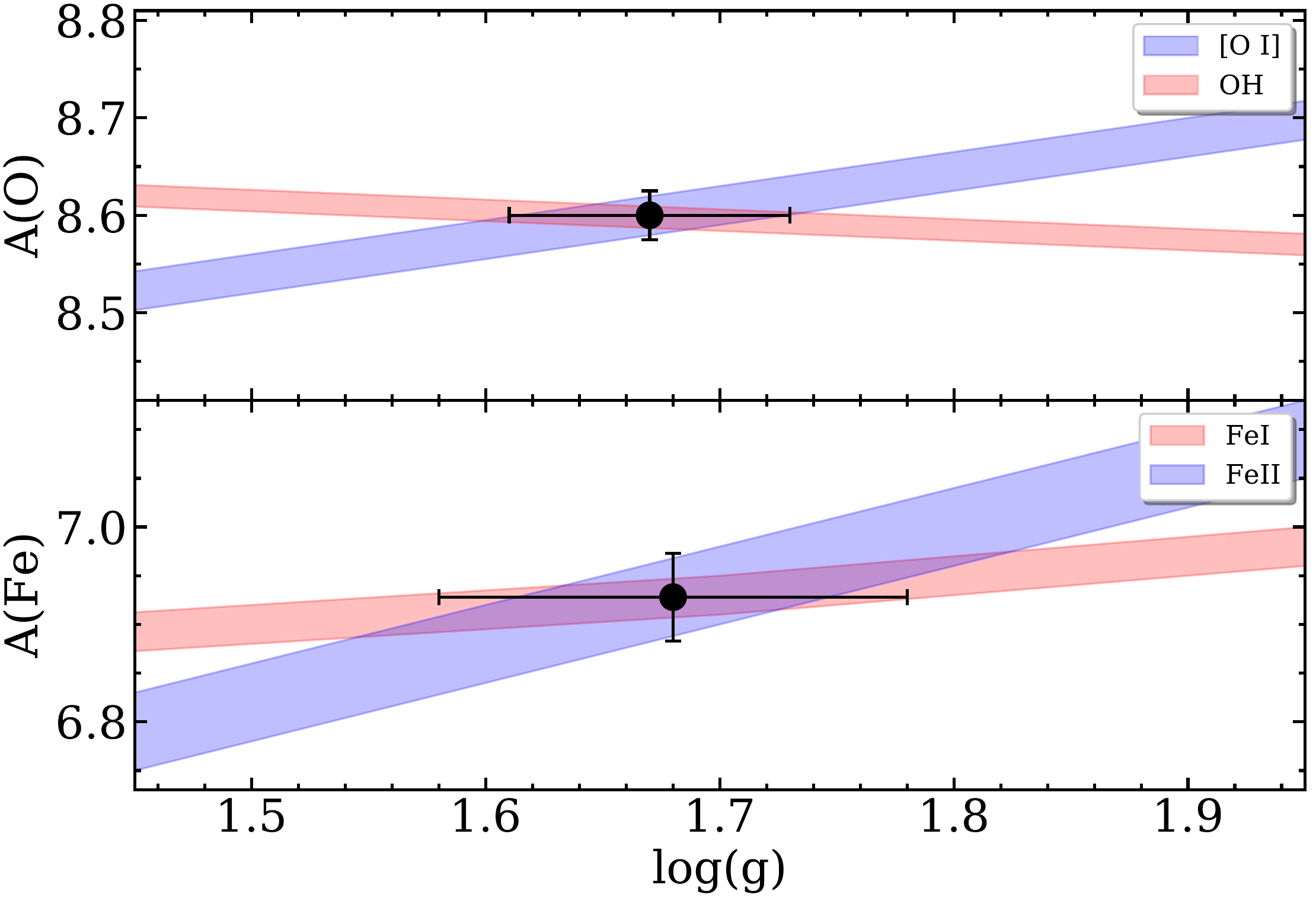}
    \caption{\textit{Top panel}: best-fit trends of Oxygen abundances from the [O I] line (blue) and from the OH lines (red) with gravity.
    The black dot marks the intersection of the two curves, and its x-coordinate provides the best-fit log(g)=1.67$\pm 0.06$. \textit{Bottom panel}: best-fit trends of Iron abundances from Fe II (shaded) and Fe I lines (red) with gravity. The black dot marks the intersection of the two curves, and its x-coordinate provides the best-fit log(g)=$1.68\pm 0.10$.
    }\label{OFe_gravi}
\end{figure}

After fixing the temperature at T$_{eff}=4283$ K, as derived from the C-thermometer, we can fine-tune the gravity by balancing the O abundances derived from the forbidden transition [O~I] at $6300.3\, \AA$, and from the numerous OH lines measurable in the NIR spectrum.
Fig.~\ref{OFe_gravi} (top panel) shows the variation in O abundance from the [O I] and OH lines in the Arcturus optical and NIR spectra, respectively, as a function of log(g). 
The intersection of the two curves occurs at log(g)$=1.67\pm 0.06$ dex. 
The quoted error of $\pm$0.06~dex corresponds to a $\pm1\sigma$ variation in the derived O abundances from [OI] and OH.

Variations of $\pm$ 33 K in temperature affect gravity by $\pm$0.10 dex, while variations of $\pm 0.05$ $km s^{-1}$ in microturbulence velocity affect log(g) by $\mp0.02$ dex. Using the standard method for inferring spectroscopic log(g), that is, minimising the difference between the iron abundances from neutral and ionised optical lines, we  obtained a very similar best-fit log(g)=$1.68\pm 0.10$ dex (see Fig.~\ref{OFe_gravi}, bottom panel).

\section{Chemical analysis}\label{chem}
The adopted stellar parameters for the chemical analysis of Arcturus are summarised in Table~\ref{param}. 
\begin{table}[!h]
\centering
\caption{\label{param} 
Stellar parameters for Arcturus}
\begin{tabular}{lll}
\hline
Paramater & Value & Error   \\
\hline
\rowcolor{lightgray}
 T$_{eff}$ & $4283$ K & $33$ K  \\
 log(g) & $1.67$ dex & $0.06$ dex  \\
\rowcolor{lightgray}
 $\xi$ & $1.60$ $km s^{-1}$ & $0.05$ $km s^{-1}$  \\
 $[Fe/H]^*$ & $-0.57$  dex & $0.01$  dex  \\
\hline\hline
\end{tabular}
\tablefoot{\\
   \tablefoottext{*}{We used the solar A(Fe)$_{\odot}$=7.50, as in \citet{Grevesse98} and \citet{Asplund09}. \\}}
\end{table}
Abundance errors from the uncertainties in the stellar parameters were 
estimated by computing elemental abundances with varying $T_{eff}$ by $\pm33$~K, log(g) by $\pm0.06$~dex, and $\xi$ by $\pm0.05$~$km s^{-1}$ (see Table~\ref{param}). 

\begin{landscape}
\begin{table}[!h]
\centering
\caption{Arcturus chemical abundances and associated measurement errors from NIR and optical lines.}\label{abutab} 
\tiny
\begin{tabular}{l|c|ccccccc|ccccccc|cccccc}
\hline
  &  &  &  &  & \textbf{NIR} &  &  &  &  &  &  & \textbf{OPT} &  &  &  &  &  & \textbf{OPT+NIR} &  &  & \\
X & Z & \#  & log(N)$^a$ & [X/H] & [X/H] & [X/Fe] & [X/Fe] & $\epsilon$ & \#  & log(N)$^a$ & [X/H] & [X/H] & [X/Fe] & [X/Fe] & $\epsilon$  & log(N)$^a$ & [X/H] & [X/H] & [X/Fe] & [X/Fe] & $\epsilon$ \\
 & & lines & & Gre98 & Aspl09 & Gre98 & Aspl09 & & lines & & Gre98 & Aspl09 & Gre98 & Aspl09 & & & Gre98 & Aspl09 & Gre98 & Aspl09 & \\
\hline
 C     & $6$ & $ 19^b$ & $7.97$ & $-0.55$ & $-0.46$ &$+0.01$ & $+0.10$ & $0.02$ & $  1$ & $8.00$ & $-0.52$ & $-0.43$ & $+0.05$ & $+0.14$ & $0.05$ & $7.97$ & $-0.55$ & $-0.46$ & $+0.02$ & $+0.11$ & $0.06$  \\
 N$^b$ & $7$ & $137$ & $7.64$ & $-0.28$ & $-0.19$ & $+0.28$ & $+0.37$ & $0.01$ & 0 & -- & -- & -- & -- & -- & -- & -- & -- & -- & -- & -- \\ 
 O$^b$ & $8$ & $ 14$ & $8.60$ & $-0.23$ & $-0.09$ & $+0.33$ & $+0.47$ & $0.02$ & $  1$ & $8.61$ & $-0.22$ & $-0.08$ & $+0.35$ & $+0.49$ & $0.02$ & $8.60$ & $-0.23$ & $-0.09$ & $+0.34$ & $+0.48$ & $0.06$ \\
 F$^b$ & $9$ & $  1$ & $4.22$ & $-0.34$ & $-0.34$ & $+0.22$ & $+0.22$ & $0.05$ & 0 & -- & -- & -- & -- & -- & -- & -- & -- & -- & -- & -- \\ 
\hline
\rowcolor{lightgray}  
 Fe I& $26$ & $179$ & $6.94$ & $-0.56$ & $-0.56$ & $+0.00$ & $+0.00$ & $0.01$ & $230$ & $6.93$ & $-0.57$ & $-0.57$ & $+0.00$ & $+0.00$ & $0.01$ & $6.93$ & $-0.57$ & $-0.57$ & $+0.00$ & $+0.00$ & $0.01$ \\ 
\rowcolor{lightgray}  
 Fe II& $26$ & $  1$ & $6.95$ & $-0.55$ & $-0.55$ & $+0.01$ & $+0.01$ & $0.04$ & $ 10$ & $6.94$ & $-0.56$ & $-0.56$ & $+0.01$ & $+0.01$ & $0.03$ & $6.94$ & $-0.56$ & $-0.56$ & $+0.01$ & $+0.01$ & $0.05$ \\
\rowcolor{lightgray}  
 V I & $23$ & $  4$ & $3.48$ & $-0.52$ & $-0.45$ & $+0.04$ & $+0.11$ & $0.07$ & $ 41$ & $3.49$ & $-0.51$ & $-0.44$ & $+0.06$ & $+0.13$ & $0.03$ & $3.49$ & $-0.51$ & $-0.44$ & $+0.06$ & $+0.13$ & $0.08$ \\ 
\rowcolor{lightgray}  
 Cr I& $24$ & $ 29$ & $5.04$ & $-0.63$ & $-0.60$ & $-0.07$ & $-0.04$ & $0.03$ & $ 33$ & $5.02$ & $-0.65$ & $-0.62$ & $-0.08$ & $-0.05$ & $0.02$ & $5.03$ & $-0.64$ & $-0.61$ & $-0.07$ & $-0.04$ & $0.04$ \\ 
\rowcolor{lightgray}  
 Mn I& $25$ & $  4$ & $4.84$ & $-0.55$ & $-0.59$ & $+0.01$ & $-0.03$ & $0.04$ & $  5$ & $4.83$ & $-0.56$ & $-0.60$ & $+0.01$ & $+0.03$ & $0.05$ & $4.83$ & $-0.56$ & $-0.60$ & $+0.01$ & $+0.03$ & $0.06$ \\ 
\rowcolor{lightgray}  
 Co I& $27$ & $  6$ & $4.44$ & $-0.48$ & $-0.55$ & $+0.08$ & $+0.01$ & $0.06$  & $ 36$ & $4.44$ & $-0.48$ & $-0.55$ & $+0.09$ & $+0.02$ & $0.02$ & $4.44$ & $-0.48$ & $-0.55$ &  $+0.09$ & $+0.02$ & $0.06$ \\
\rowcolor{lightgray}  
 Ni I& $28$ & $ 31$ & $5.73$ & $-0.52$ & $-0.49$ & $+0.04$ & $+0.07$ & $0.02$  & $ 42$ & $5.69$ & $-0.56$ & $-0.53$ & $+0.01$ & $+0.04$ & $0.01$ & $5.71$ & $-0.54$ & $-0.51$ & $+0.03$ & $+0.06$ & $0.02$ \\
\rowcolor{lightgray}  
 Cu I& $29$ & $  1$ & $3.83$ & $-0.38$ & $-0.36$ & $+0.18$ & $+0.20$ & $0.04$  & $  1$ & $3.81$ & $-0.40$ & $-0.38$ & $+0.17$ & $+0.19$ & $0.03$ & $3.82$ & $-0.39$ & $-0.37$ & $+0.18$ & $+0.20$ & $0.05$ \\ 
\rowcolor{lightgray}  
 Zn I & $30$ & $  2$ & $4.19$ & $-0.41$ & $-0.37$ & $+0.15$ & $+0.19$ & $0.07$ & $  1$ & $4.19$ & $-0.41$ & $-0.37$ & $+0.16$ & $+0.20$ & $0.06$ & $4.19$ & $-0.41$ & $-0.37$ & $+0.16$ & $+0.20$ & $0.09$ \\
\hline
 Mg I& $12$ & $ 11$ & $7.33$ & $-0.25$ & $-0.27$ & $+0.31$ & $+0.29$ & $0.02$ & $  7$ & $7.36$ & $-0.23$ & $-0.25$ & $+0.34$ & $+0.32$ & $0.02$ & $7.34$ & $-0.24$ & $-0.26$ & $+0.33$ & $+0.31$ & $0.03$ \\
 Si I& $14$ & $ 58$ & $7.26$ & $-0.29$ & $-0.25$ & $+0.27$ & $+0.31$ & $0.03$ & $ 40$ & $7.26$ & $-0.29$ & $-0.25$ & $+0.28$ & $+0.32$ & $0.02$ & $7.26$ & $-0.29$ & $-0.25$ & $+0.28$ & $+0.32$ & $0.04$ \\
 S I& $16$ & $ 11$ & $6.90$ & $-0.43$ & $-0.22$ & $+0.13$ & $+0.34$ & $0.05$ & $  3$ & $6.91$ & $-0.42$ & $-0.21$ & $+0.15$ & $+0.36$ & $0.05$ & $6.90$ & $-0.43$ & $-0.22$ & $+0.14$ & $+0.35$ & $0.07$ \\
 Ca I& $20$ & $ 24$ & $5.84$ & $-0.52$ & $-0.50$ & $+0.04$ & $+0.06$ & $0.02$ & $ 22$ & $5.80$ & $-0.56$ & $-0.54$ & $+0.01$ & $+0.03$ & $0.02$ & $5.82$ & $-0.54$ & $-0.52$ & $+0.03$ & $+0.05$ & $0.03$ \\
 Ti I & $22$ & $ 57$ & $4.59$ & $-0.43$ & $-0.36$ & $+0.13$ & $+0.20$ & $0.02$ & $112$ & $4.55$ & $-0.47$ & $-0.40$ & $+0.10$ & $+0.17$ & $0.01$ & $4.56$ & $-0.46$ & $-0.39$ & $+0.11$ & $+0.18$ & $0.02$ \\
 Ti II& $22$ & 0 & -- & -- & -- & -- & -- & -- & $ 21$ & $4.61$ & $-0.41$ & $-0.34$ & $+0.16$ & $+0.23$ & $0.03$ & -- & -- & -- & -- & -- & --  \\
\hline
\rowcolor{lightgray}  
 Na I& $11$ & $  3$ & $5.78$ & $-0.55$ & $-0.46$ & $+0.01$ & $+0.10$ & $0.04$ & $  3$ & $5.76$ & $-0.57$ & $-0.48$ & $+0.00$ & $+0.09$ & $0.03$ & $5.77$ & $-0.56$ & $-0.47$ & $+0.01$ & $+0.10$ & $0.04$ \\
\rowcolor{lightgray}  
 Al I& $13$ & $  4$ & $6.14$ & $-0.33$ & $-0.31$ & $+0.23$ & $+0.25$ & $0.04$ & $ 10$ & $6.15$ & $-0.32$ & $-0.30$ & $+0.25$ & $+0.27$ & $0.02$ & $6.15$ & $-0.32$ & $-0.30$ & $+0.25$ & $+0.27$ & $0.04$ \\
\rowcolor{lightgray}  
 P I& $15$ & $  2$ & $5.14$ & $-0.31$ & $-0.27$ & $+0.25$ & $+0.29$ & $0.06$ & 0 & -- & -- & -- & -- & -- & -- & -- & -- & -- & -- & -- & -- \\ 
\rowcolor{lightgray}  
 K I & $19$ & $  6$ & $4.81$ & $-0.31$ & $-0.22$ & $+0.25$ & $+0.34$ & $0.07$ & $  2$ & $4.80$ & $-0.32$ & $-0.23$ & $+0.25$ & $+0.34$ & $0.08$ & $4.81$ & $-0.31$ & $-0.22$ & $+0.26$ & $+0.35$ & $0.11$ \\
\rowcolor{lightgray}  
 Sc I & $21$ & $  2$ & $2.66$ & $-0.51$ & $-0.49$ & $+0.05$ & $+0.07$ & $0.04$ & $ 20$ & $2.65$ & $-0.52$ & $-0.50$ & $+0.05$ & $0.07$ & $0.03$ & $2.65$ & $-0.52$ & $-0.50$ & $+0.05$ & $+0.07$ & $0.05$ \\
\rowcolor{lightgray}  
 Sc II& $21$ & 0 & -- & -- & -- & -- & -- & -- & $ 18$ & $2.71$ & $-0.46$ & $-0.44$ & $+0.11$ & $+0.13$ & $0.04$ & -- & -- & -- & -- & -- & -- \\
\hline
 Y I & $39$ & 0 & -- & -- & -- & -- & -- & -- & $  1$ & $1.51$ & $-0.73$ & $-0.70$ & $-0.16$ & $-0.13$ & $0.10$ & -- & -- & -- & -- & -- & -- \\  
 Y II& $39$ & $ 1$  & $1.51$ & $-0.73$ & $-0.70$ & $-0.17$ & $-0.14$ & $0.08$ & $ 5$ & $1.56$ & $-0.68$ & $-0.65$ & $-0.11$ & $-0.08$ & $0.06$ & $1.54$ & $-0.70$ & $-0.67$ & $+0.13$ & $+0.10$ & $0.10$ \\ 
 Ce II& $58$ & $ 6$  & $1.06$ & $-0.52$ & $-0.52$ & $+0.04$ & $+0.04$ & $0.04$ & $ 4$ & $1.04$ & $-0.54$ & $-0.54$ & $+0.03$ & $+0.03$ & $0.05$ & $1.05$ & $-0.53$ & $-0.53$ & $+0.04$ & $+0.04$ & $0.06$ \\ 
 Nd II& $60$ & $ 1$  & $0.93$ & $-0.57$ & $-0.65$ & $-0.01$ & $-0.09$ & $0.06$ & $ 19$ & $0.90$ & $-0.60$ & $-0.68$ & $-0.03$ & $-0.11$ & $0.03$ & $0.90$ & $-0.60$ & $-0.68$ & $-0.03$ & $-0.11$ & $0.07$ \\ 
 Dy II& $66$ & $  1$ & $0.70$ & $-0.44$ & $-0.40$ & $+0.12$ & $+0.16$ & $0.05$ & $ 1$ & $0.68$ & $-0.46$ & $-0.42$ & $+0.11$ & $+0.15$ & $0.05$ & $0.69$ & $-0.45$ & $-0.41$ & $+0.12$ & $+0.16$ &  $0.07$ \\   
\hline\hline
\end{tabular}
   \tablefoot{\\
   \tablefoottext{$a$}{log(N)=log(N$_{X}$/N$_H$)+12.}\\
   \tablefoottext{$b$}{C abundance has been derived by combining the measurement of 5 C~I atomic lines and 14 CO molecular lines. N, O, and F abundances have been derived from the measurement of molecular lines only.}\\
   }
\end{table}
\end{landscape}

On average, these errors amount to a few hundredths of a dex at most.
The only notable exception is the higher sensitivity of OH and HF lines to the effective temperature: a variation of $\pm$33~K indeed implies an error in the derivation of oxygen and fluorine abundances of $\pm$0.06 and $\pm$0.07 dex, respectively.

In the computation of these errors we did not include the interdependence between the  C, N, and O abundances that contribute to the formation of the measured molecular lines nor the effect of the abundances of the main electron donors on those derived from ionised species. We estimate that these effects normally yield errors below 0.1 dex \citep[see also e.g.][]{Ryde09}.

Measurement errors include uncertainty in the continuum positioning and photon noise. For elements with more than two measurable lines, we computed the dispersion around the mean abundance, while for those with one or two measurable lines, we computed the dispersion from a Monte Carlo simulation,  taking into account an error in the measured EW of $\approx\pm FWHM/\text{signal-to-noise ratio} $ for a line FWHM sampled with 2-3 pixels.\\
The measurement errors $\epsilon$ quoted in Table~\ref{abutab} are the $\sigma$ dispersion divided by the square root of the number of lines.\\
In the course of the chemical analysis we checked a few problematic lines against possible NLTE effects using the online web tool  \url{http://nlte.mpia.de/gui-siuAC_secE.php}, \citealt{NLTE_MPIA}. \\
Abundances and corresponding measurement errors for all the sampled chemical elements in the log(N$_X$/N$_H$)+12 and in the [X/H] solar scales, adopting  as solar reference both \citet{Grevesse98} (Gre98) and \citet{Asplund09} (Aspl09), are listed in Table~\ref{abutab}.\\
In Fig.~\ref{abu_IRopt} we compare the derived abundances from NIR and optical lines for all the measured elements.

\subsection{CNO and fluorine}
\label{cno}

In our chemical analysis of Arcturus we first computed the abundances of CNO and then those of the other elements. CNO are the most abundant metals, and in red giant and supergiant star spectra the many molecular CO, CN, and OH lines are the most important potential contaminants.
Following \cite{Ryde09} and \cite{Smith13}, 
we adopted an iterative method to derive CNO abundances in order to consider the interplay among these three elements in setting the molecular equilibrium. 
The resulting abundances are listed in first three lines of Table~\ref{abutab}.

Most interestingly, the  carbon abundance derived from the $\Delta v$=3 CO band-heads visible in the H band is 8.02 and equal within the errors with the values derived from atomic C~I lines and isolated CO roto-vibrational transitions.\\
This result indicates that the $\Delta v$=3 CO band-heads could be effectively used to measure carbon abundances in cold stars where the single CO lines are severely blended or too weak for a reliable abundance analysis.\\
Following a similar approach, we estimated the abundance of $^{13}$C using two $\Delta v$=3 and three $\Delta v$=2 band-heads of $^{13}$CO. This yielded a $^{12}$C/$^{13}$C isotopic abundance ratio of $7\pm1$, consistent with the effects of the second dredge-up coupled to some additional mixing in low-mass giants \citep[see e.g.][]{Charbonnel&Lagarde10}.

The fluorine abundance was derived from one HF molecular line at $23358.33 \, \AA$ in the K band using the transition parameters of \citealt{Jonsson14}. 

\subsection{Iron-peak elements}\label{iron}

\begin{figure}[h!]
    \centering
    \includegraphics[width=9cm,height=4.cm]{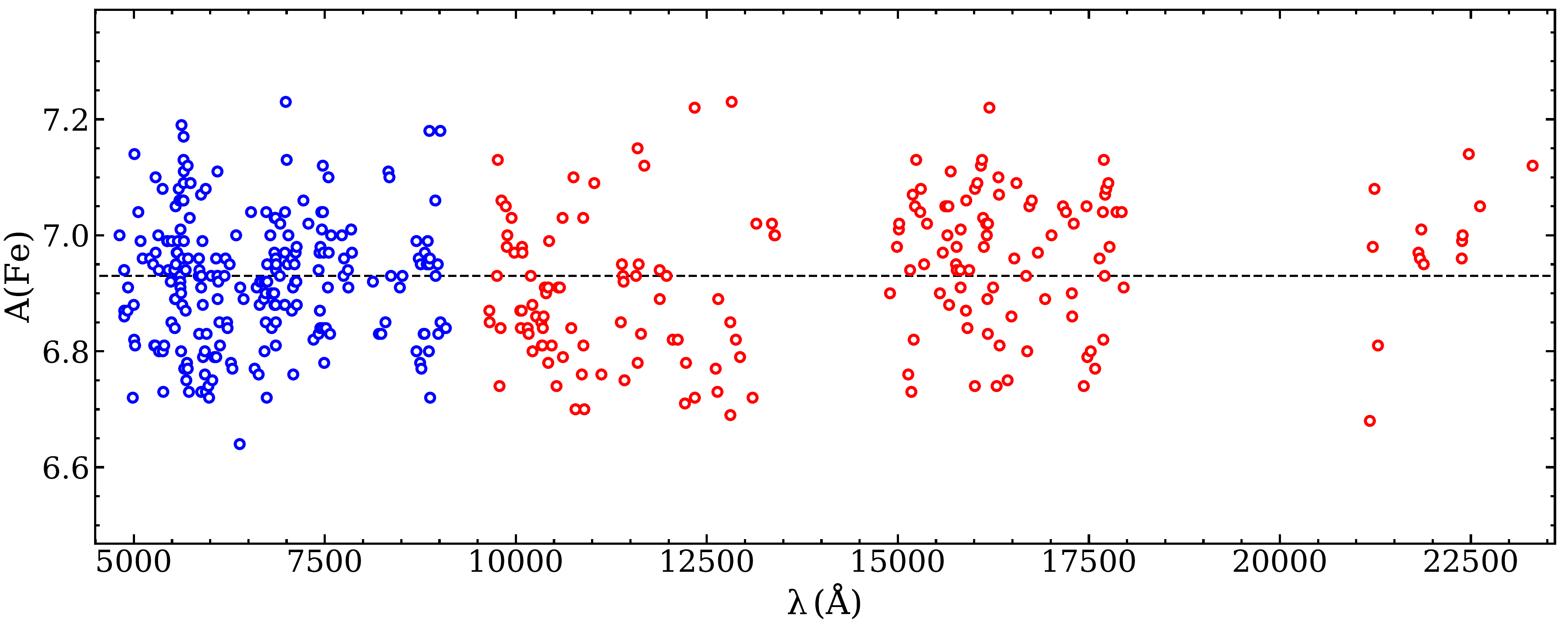}
    \caption{Iron abundances for all the measured optical and NIR lines as a function of the line central wavelength. The dashed line marks the average abundance.}
    \label{Fe_wl}
\end{figure}

The iron abundance was derived for more than 400 lines of Fe~I and a dozen Fe~II lines.
Fully consistent values were obtained from all these line sets, with a
$\sigma$ dispersion of about 0.1 dex. 
In Fig.~\ref{Fe_wl} we show the inferred iron abundances from the measured  iron lines 
over the full 4800-24500 spectral range as a function  of the line central wavelength. 
We adopted the average value of $6.93\pm 0.01$ (see Table~\ref{abutab}) as the reference iron abundance of Arcturus.
However, it is interesting to note that many of the lines in the YJ bands provide abundances that are systematically lower (on average by 0.04 dex) than the reference abundance. We inspected some of these lines for possible NLTE effects and found indeed that some positive corrections of $\le$0.1 dex should be applied to their LTE abundances.

For the other iron-peak elements, we were also able measure both optical and NIR lines and derived suitable abundances. In particular, for chromium and nickel, we used a few dozen lines both in the optical and NIR range. For vanadium we used a few dozen lines in the optical with hyperfine splitting (hereafter HFS) and only a few in the NIR. 

For manganese we used the few measurable lines both in the optical and NIR. In particular, in the NIR we used the J-band lines at $13218.49 \, \AA$ and $13415.64 \, \AA$ that need an NLTE correction of $\Delta A(Mn)=-0.16$ and $\Delta A(Mn)=-0.04$, respectively, and two lines in the H band with HFS. 
For cobalt we used a few dozen optical lines with HFS and a few NIR lines. The latter show a larger scatter that can be explained with small NLTE effects. For example, we applied an NLTE correction of $+0.06$ dex to the LTE abundance of the $16757.64 \, \AA$ line.
For copper we were only able use one optical and one NIR line. The NIR line at $16005.75 \, \AA$ has HFS and needs to be used with caution because it might be blended. 
For zinc we used one line in the optical and two in the NIR. The NIR line at $11054.28 \, \AA$ is partially blended with CN, but the line at $13053.63 \, \AA$ is free of contamination. 

\begin{figure*}[!h]   \label{abu_IRopt}
    \centering
    \includegraphics[scale=0.475]{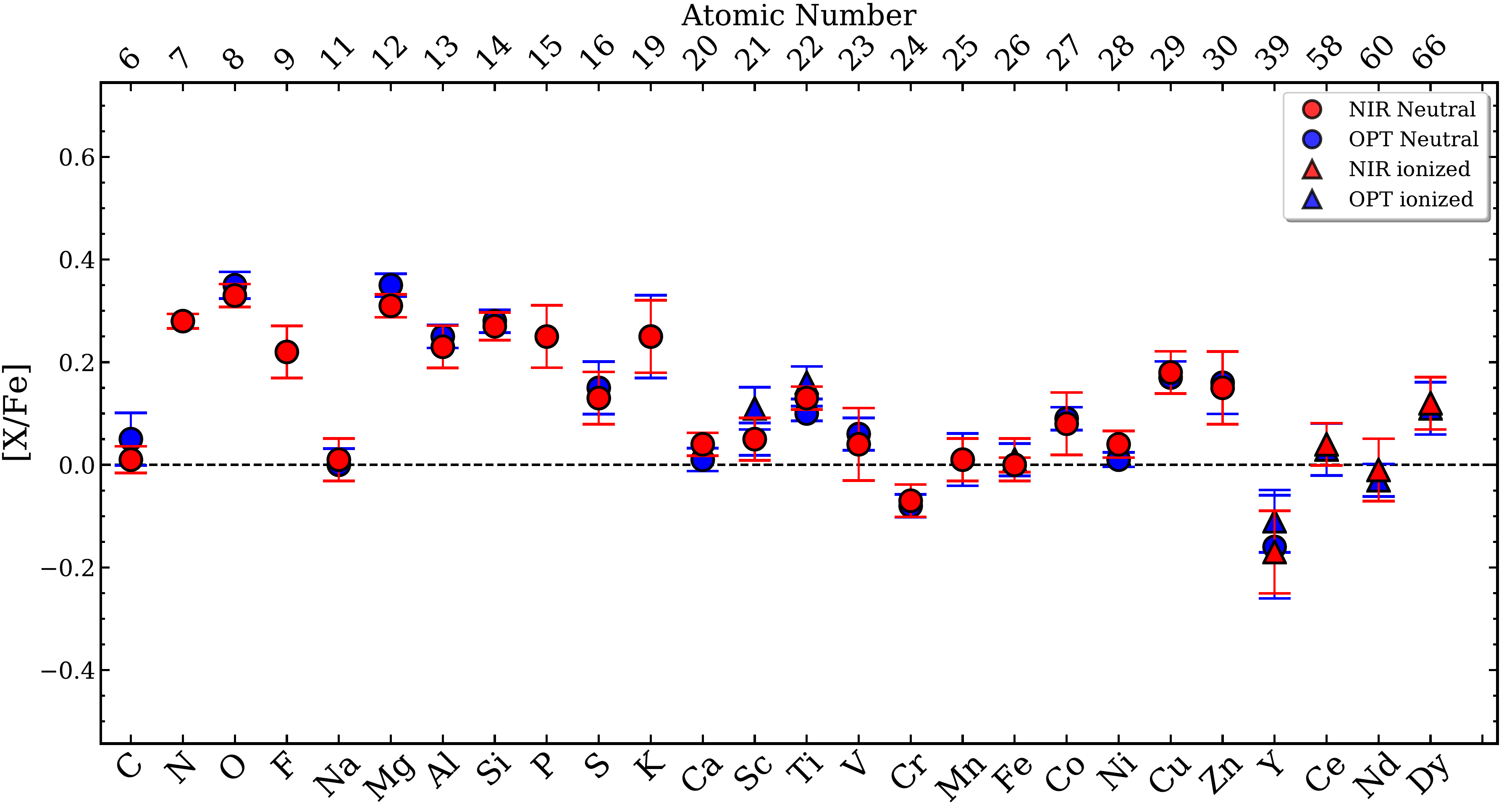}
    \caption{Derived [X/Fe] chemical abundances for Arcturus from the GIANO-B NIR spectrum (red symbols) and UVES optical spectra (blue symbols). The circles indicate the neutral species, and the triangles indicate the ionised species. Error bars are from Table~\ref{abutab}.}
\end{figure*}

Altogether, the iron-peak elements show fully consistent optical and NIR abundances. They homogeneously scale as iron, with the possible exception of copper and zinc, which are slightly enhanced. 

\subsection{$\alpha$-capture elements}
Dozens of unblended lines of Si~I, Ca~I, Ti~I, Mg~I, and S~I are available in the NIR spectrum of Arcturus for an abundance analysis.
The NIR Mg~I and S~I lines are known to experience NLTE effects \citep{Zhang17,Takeda16sulfurzinc}. However, at the metallicity, temperature, and gravity of Arcturus, the corrections are negligible.
Sulphur also shows a forbidden line [S I] at $10821 \, \AA$ that provides a fully consistent abundance with the one derived from the selected S~I lines. This further proves that NLTE effects are negligible.

The inferred NIR abundances for these alpha elements are fully consistent with the optical ones, as detailed in Table~\ref{abutab} and shown in Fig.~\ref{abu_IRopt}. \\
The derived abundances of Mg and Si, and to a lesser extent, of Ti, S, and Ca, suggest some [$\alpha$/Fe] enhancement, as for oxygen. This is typical of thick-disk stars.

\subsection{Z-odd elements}
A few optical and NIR lines of Na, K, and Sc can be safely used to derive reliable abundances. We found consistent optical and NIR solar-scaled abundances of Na and Sc and some enhanced K abundance. However, the K lines can show significant NLTE effects with negative corrections to the LTE abundances \citep[see e.g.][]{zhang06,osorio20}, thus implying a lower [K/Fe] relative abundance, about solar or even subsolar scaled.
Two NIR lines of phosphorus at $10529.52 \, \AA$ and $10581.58 \, \AA$ are also measurable, giving an abundance of $5.14\pm0.08$. 
A third line at  $10596.90 \, \AA$  gives a  unexpected higher abundance  \citep[see also][]{Maas17}, probably blended because its profile is clearly asymmetric. We therefore rejected it.

For aluminium, ten optical lines and four NIR lines at $10782.05 \, \AA$, $10768.37 \, \AA$,  $10872.97 \, \AA,$ and  $10891.77 \, \AA$ in the Y band with small (if any) NLTE corrections provide homogeneous abundances that are higher by almost a factor of two than the solar-scaled value. For the NIR lines we used log(gf) from NIST, which is slightly different from the lines adopted in VALD3. 

The strong lines at $13123.41 \, \AA$ and $13150.75 \, \AA$ and the K-band line at at $21163.76 \, \AA$ have HFS, show significant NLTE effects \citep{Nordlander17}, and require a negative abundance correction of $0.25-0.30$ dex. \\
Although when these lines are corrected for NLTE, they provide Al abundances that are reasonably consistent with those of the Y band and optical lines, we did not use them.

The three strong lines at $16718.96 \, \AA$,  $16750.56 \, \AA,$ and  $16763.36 \, \AA$ also show NLTE effects and have HFS. The line at $16750.56 \, \AA$ also has strong and blended wings.\\
In Arcturus-like stars, the abundances derived from these lines can be quite uncertain, therefore we did not use them (see Sect.~\ref{chromospheric_activity}). 

\subsection{Neutron-capture elements}

We measured NIR lines for five neutron-capture elements: yttrium (mostly an s-process element), cerium (an s-process element), neodimium (mostly an s-process element), and dysprosium (an r-process element). The NIR Ce~II and Nd~II lines were  identified for the first time by \citet{Cunha17} and \citet{Hasselquist16}, respectively.

One neutral and five ionised lines of yttrium were measured in the optical, but only one ionised line is measured in the NIR Y band \citep[see also][]{Mat20}. We find that the Y abundance is slightly depleted with respect to the solar scale value, in agreement with the disk chemistry at the Arcturus metallicity \citep[see e.g.][]{Reddy06,bensby14}.

We finally used a few optical and NIR-ionised lines of cerium, neodymium, and dysprosium, and we derived about solar-scaled Ce and Nd abundances and slightly enhanced Dy with respect to the solar-scaled value.

\begin{figure}[!h]
    \centering
    \includegraphics[width=9cm,height=5cm]{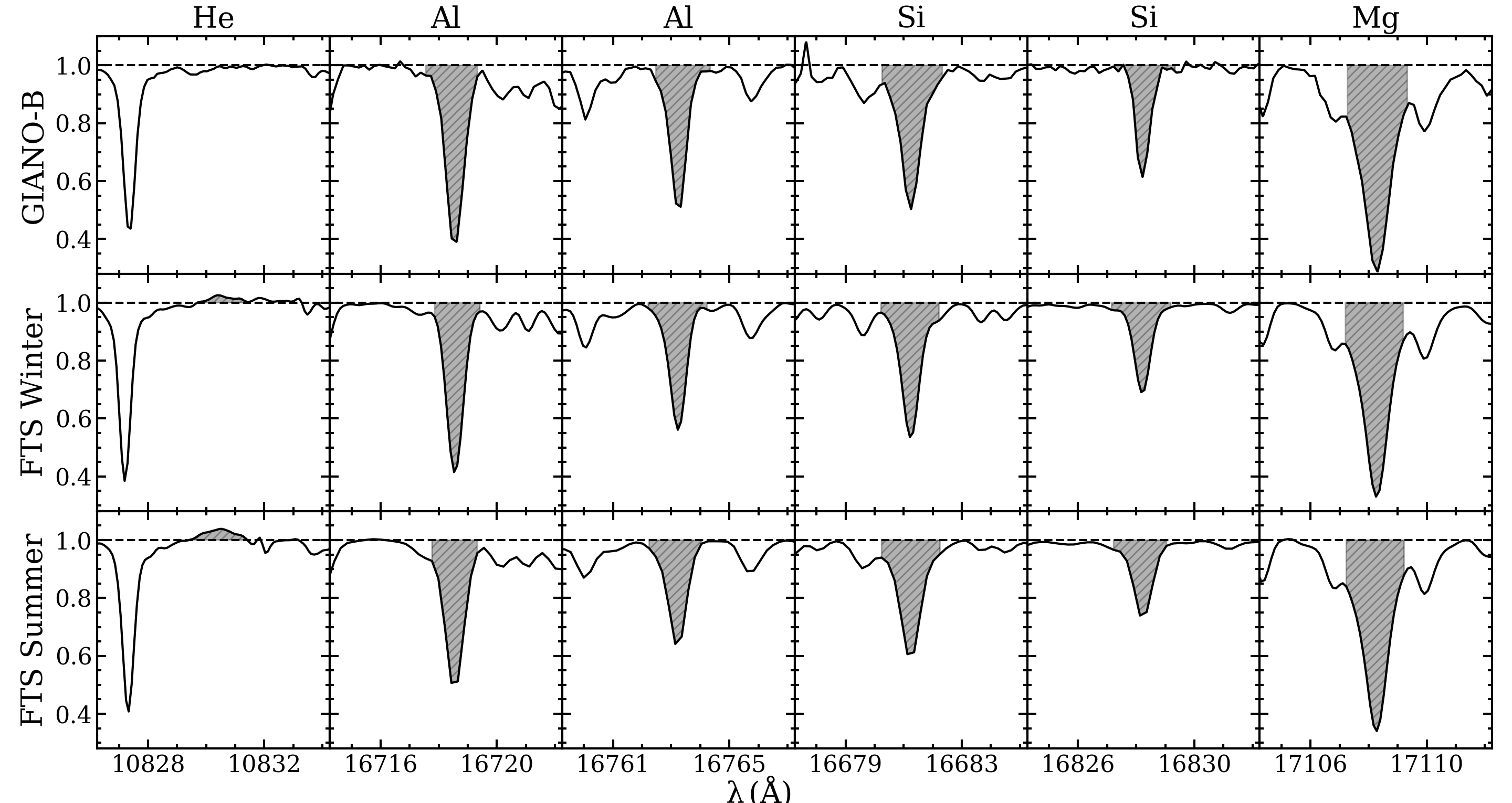}
    \caption{Chromospheric He~I line in the Y band and a few strong photospheric  Al, Si, and Mg lines in the in the GIANO-B spectrum (top panel) and in the winter, January (middle panel) and summer, June (bottom panel) FTS spectra by \citet{Hinkle05}.}
    \label{He_Al}
\end{figure}

\subsection{Chromospheric activity}\label{chromospheric_activity}

We realised that some strong lines in the GIANO-B spectrum are 
deeper than the corresponding lines in the FTS winter and summer spectra of Arcturus by \citet{Hinkle00}. A few examples are shown in Fig.~\ref{He_Al}.

Chromospheric activity can fill the core of 
strong lines and mimic shallower absorptions \citep[e.g.][]{Shcherbakov96}.
We therefore wondered whether a variation for the chromospheric activity in Arcturus might cause the different line depth in the GIANO-B and FTS spectra.  

For this purpose, we used the He~I line at $10830 \, \AA,$ which is a good indicator of chromospheric activity \citep{Danks85}.
As shown in Fig.~\ref{He_Al}, when the winter and especially the summer FTS spectra were acquired, chromospheric activity was higher, as suggested by some He~I emission and  shallower photospheric lines, while when the GIANO-B spectrum was acquired, the activity was low, without He~I emission and with deeper photospheric lines. Because strong lines can be problematic also because they might be weakened by this chromospheric activity, they should be used with great caution for an abundance analysis. We excluded these lines from our abundance analysis. 

\begin{figure*}[!h]
    \centering
    \includegraphics[scale=0.477]{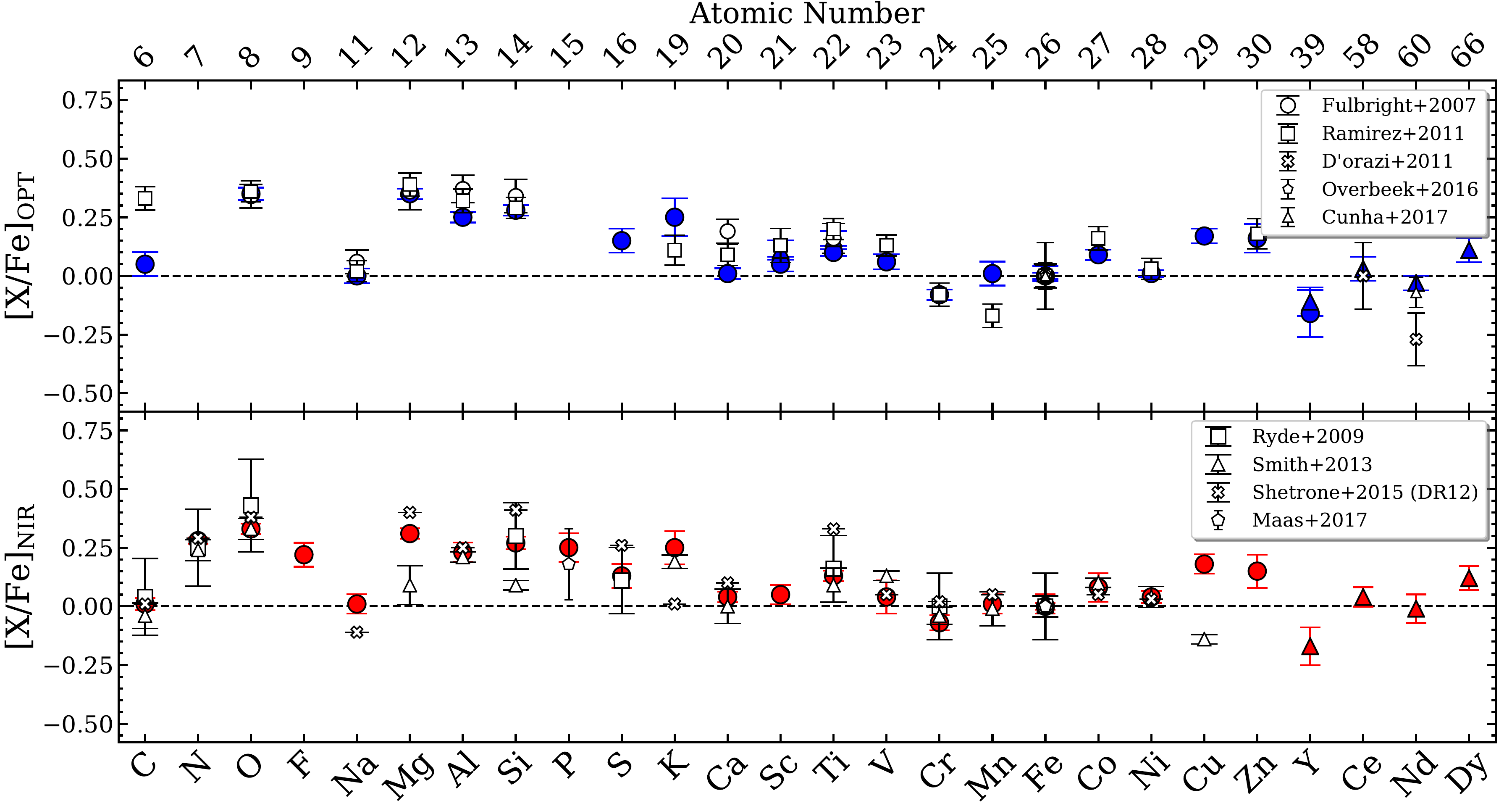}
    \caption{Derived [X/Fe] chemical abundances for Arcturus from some optical (top panel) and NIR (bottom panel) studies quoted in Table~\ref{reftab} and in Sect.~\ref{intro}.
    Blue symbols are our abundances from the optical UVES spectra, and red symbols our abundances from the NIR GIANO-B spectrum. Blue and red dots refer to neutral species, and blue and red triangles to ionised species.}
    \label{abu_ref}
\end{figure*}

\section{Discussion and conclusions}

Detailed high-resolution optical and NIR spectroscopy of stellar calibrators 
is fundamental for defining optimal diagnostics for atmospheric parameters 
and chemical analysis of stars and stellar populations with different ages, metallicities, and 
evolutionary properties.

While diagnostic tools from high-resolution optical spectroscopy are well established and have been calibrated for a long time, those from NIR spectroscopy have been begun to be explored only recently with the new generation of NIR echelle spectrographs, whose 
 performances are suitable for such quantitative studies.

We used Arcturus as a laboratory to explore optical and  NIR spectroscopic diagnostics for chemical analyses over the full spectral range from 4800 to 24500. We then provided a comprehensive and self-consistent determination of the stellar parameters and chemical abundances of Arcturus.

The value of this combined optical and NIR study is multifold and is summarised below.

{\it i})~The study maximises the set of diagnostic lines so that almost all the chemical elements of interest can be sampled from a statistically significant number of lines for most of them. 

{\it ii})~The study enables sampling lines of a given species at different wavelengths, which extends the range of excitation potentials and transition probabilities for a better understanding of the physics 
of line formation and the modelling of the observed spectrum.

{\it iii})~The study drives the analysis towards a physical self-consistent solution over the entire spectrum of the degeneracy 
problem among stellar parameters and chemical abundances.

Taking advantage of our optical and NIR analysis, {\it i})~we were able to set an optimal value for the microturbulence velocity that works over the full spectral range from $4800$ to $24500 \, \AA,$ and {\it ii})~we were able to define a new spectroscopic thermometer and new spectroscopic gravitometer for cool giants, based on  atomic and NIR molecular diagnostics of carbon and oxygen abundances, as detailed in Sect.~\ref{param} and Figs. \ref{Ctermo3} and \ref{OFe_gravi}. 
Using these diagnostic tools, we infer a temperature and gravity for Arcturus that are
fully consistent with photometric estimates and with the values obtained from the standard tools used in optical spectroscopy (see Figs. \ref{Fe_vs_chi} and \ref{OFe_gravi}).

As discussed in Sect.~\ref{chem} and shown in Fig.~\ref{abu_IRopt}, we find fully consistent optical and NIR abundances for all elements we analysed. This demonstrates that {\it i})~the current generation of NIR echelle spectrographs is fully adequate to deliver high-quality data for quantitative spectroscopy as in the optical, and {\it ii})~the available atomic and molecular data for the NIR lines are generally accurate enough for a reliable chemical abundance analysis.

Carbon, sodium, potassium, and iron-peak elements (with the exception of copper and zinc, which are slightly higher) are consistent with solar-scaled values, with abundances between one-fourth and one-third solar.
Nitrogen, oxygen, fluorine, and alpha elements (with the only exception of Ca, which is about solar-scaled), are higher  by a few tenths of a dex than the corresponding solar-scaled values. The values inferred for [F/O] and [F/Fe] agree with literature values for thick-disk giant stars of similar metallicity \citep[see e.g.][their figure 2]{Grisoni20}.
Among the neutron-capture elements, cerium and neodymium are about solar-scaled, dysprosium is slightly higher, and Yttrium is slightly lower than the corresponding solar-scaled values.

Our inferred stellar parameters and chemical abundances are normally fully consistent with those obtained in previous optical or NIR studies, as shown in Fig.~\ref{abu_ref}. We stress here that abundance differences of a few hundredths dex in different studies is intrinsic to the analysis process because different studies may use different codes, model atmospheres, and/or line lists and transition probabilities as well as some different assumptions for the stellar parameters. \\
In particular, when we compare our optical abundances with those obtained by \citet{Ramirez-Prieto11}, we found some notable discrepancies only for C and Mn abundances. They determined the C abundance from four C I lines. In our analysis we rejected these lines because the two at $9078 \, \AA$ and $9111 \, \AA$ are affected by NLTE and the other two at $8335 \, \AA$ and  $5380 \, \AA$ are blended.  We thus used only the forbidden line at $8727 \, \AA$.
As discussed in Sect.~\ref{subsec:C-thermometer}, our [C I] abundance is fully consistent with the C abundance derived from CO and C I lines in the H band.\\
Regarding Mn, the authors mostly measured blue lines that are absent from our UVES spectrum, while we measured lines in the red part of the optical spectrum. As discussed in Sect.~\ref{iron}, our optical Mn abundance is fully consistent with our NIR estimates and the values obtained in other NIR studies.

It is also interesting to compare our results on stellar parameters and iron abundances with those obtained by the NIR studies of \citet{Kondo19} in the in the $9300-13100 \, \AA$ spectral range and of \citet{Smith13} in the $15000-17000 \, \AA$ spectral range, at about half the spectral resolution of our study. \\
\citet{Kondo19} used a standard temperature and gravity, but a microturbulence velocity of 1.2 $km s^{-1}$, which is lower than any previous study and also lower than our adopted value of 1.6 $km s^{-1}$. They also used two different line lists, VALD3 and the  list by \citet{MB99} (hereafter MB99).
Our list has 48 lines in common with the \citet{Kondo19} VALD3 line list. 
Our and the \citet{Kondo19} VALD3 abundances are similar, although we adopted a 0.4 $km s^{-1}$ higher microturbulence velocity.\\
The other 20 lines in the \citet{Kondo19} VALD3 line list have not been used in our analysis because they are blended or contaminated by nearby strong photospheric or telluric lines. These rejected lines give an average  abundances that differs by more than 0.1 dex and has a significantly larger (>0.2 dex) dispersion. \\
We also used 30 lines in the YJ bands that are not in the VALD3 \citet{Kondo19} line list. On average, they provide abundances that are a few hundredths dex higher than those from the the lines in common.
As a result, our average  VALD3 iron abundance from our selection of  YJ band lines is 0.08 dex higher than in \citet{Kondo19} and only a few hundredths dex lower than our reference abundance from the full set of optical and NIR lines. As mentioned in  Sect.~\ref{iron}, the correction for NLTE effects can significantly mitigate if not solve the problem. \\
When we use the MB99 astrophysical $log(gf)$, which are given for lines in the $10,000-18,000 \, \AA$ range and are on average lower by 0.2-0.3 dex than those in VALD3, we find corresponding larger abundances. \\
When compared to the iron abundances obtained in previous optical studies (see Table~\ref{reftab}) and also in the present one, the abundances from the MB99 $log(gf)$ are in excess by more than 1$\sigma$ in the YJ band and more than 2$\sigma$ in the H band. When we also use the lower microturbulence velocity of \citet{Kondo19}, we obtain even larger and unlikely iron abundances.\\
\citet{Smith13} used a microturbulence velocity of 1.85 $km s^{-1}$ and nine H-band lines with astrophysical $log(gf)$ calibrated on the Sun and Arcturus IR FTS spectra by \citet{Liv91} and \citet{Hinkle05}, respectively. Their  $log(gf)$ values are somewhat in between those of VALD3 and MB99.
In our analysis we used 84 H band lines. Five out of the nine lines listed by \citet{Smith13} are in common with our sample. The other four lines in the  \citet{Smith13} list have been rejected because they are problematic (i.e. partially blended and/or with strong wings). Our and their average abundances from the five lines in common are very similar. The lower $log(gf)$ values used by \citet{Smith13} are somewhat compensated for by their slightly higher microturbulence velocity. The slightly revised iron abundance in \citet{Shetrone15} is practically coincident with our estimate.

The problem of the imperfect modelling of a line is highly degenerate. A given variation in the abundance from a given line can be obtained by modifying its log(gf), but also by slightly varying the stellar parameters and/or by using different prescriptions for the damping, HFS, NLTE corrections, etc. This degeneracy \citep[see also e.g.][]{Takeda92} cannot be easily removed. \\
The astrophysical calibration of the log(gf) is becoming very popular. However, this calibration  is model dependent (i.e. it depends on the adopted model atmospheres, spectral code, lines, etc.), and it also depends on the calibrator itself, that is, on the selected star and observed spectrum as well as on the adopted stellar parameters. None of the proposed astrophysically calibrated log(gf) can therefore be safely adopted in studies that use recipes and tools for chemical analyses that are different from those used for the astrophysical calibration. \\
Our combined optical and NIR analysis shows that it is not necessary to systematically tune the log(gf) of the NIR lines in the VALD3 database to obtain reliable abundances if the appropriate set of lines is chosen and self-consistent stellar parameters are derived.
\newline

{\it Acknowledgements.} 
We thank the anonymous referee for his/her detailed report and useful suggestions. C. Fanelli would like to thank A. Minelli for useful discussions. We acknowledge the support by INAF/Frontiera through the "Progetti Premiali" funding scheme of the Italian Ministry of Education, University, and Research. We acknowledge support from the project Light-on-Dark granted by MIUR through PRIN2017-000000 contract and support from the mainstream project SC3K - Star clusters in the inner 3 kpc funded by INAF.

\newpage

\bibliographystyle{aa}
\bibliography{mybib}
\appendix
\section{Arcturus atlas}
\begin{landscape}
\clearpage 
\begin{figure}
    \includegraphics[width=25cm,height=15.5cm]{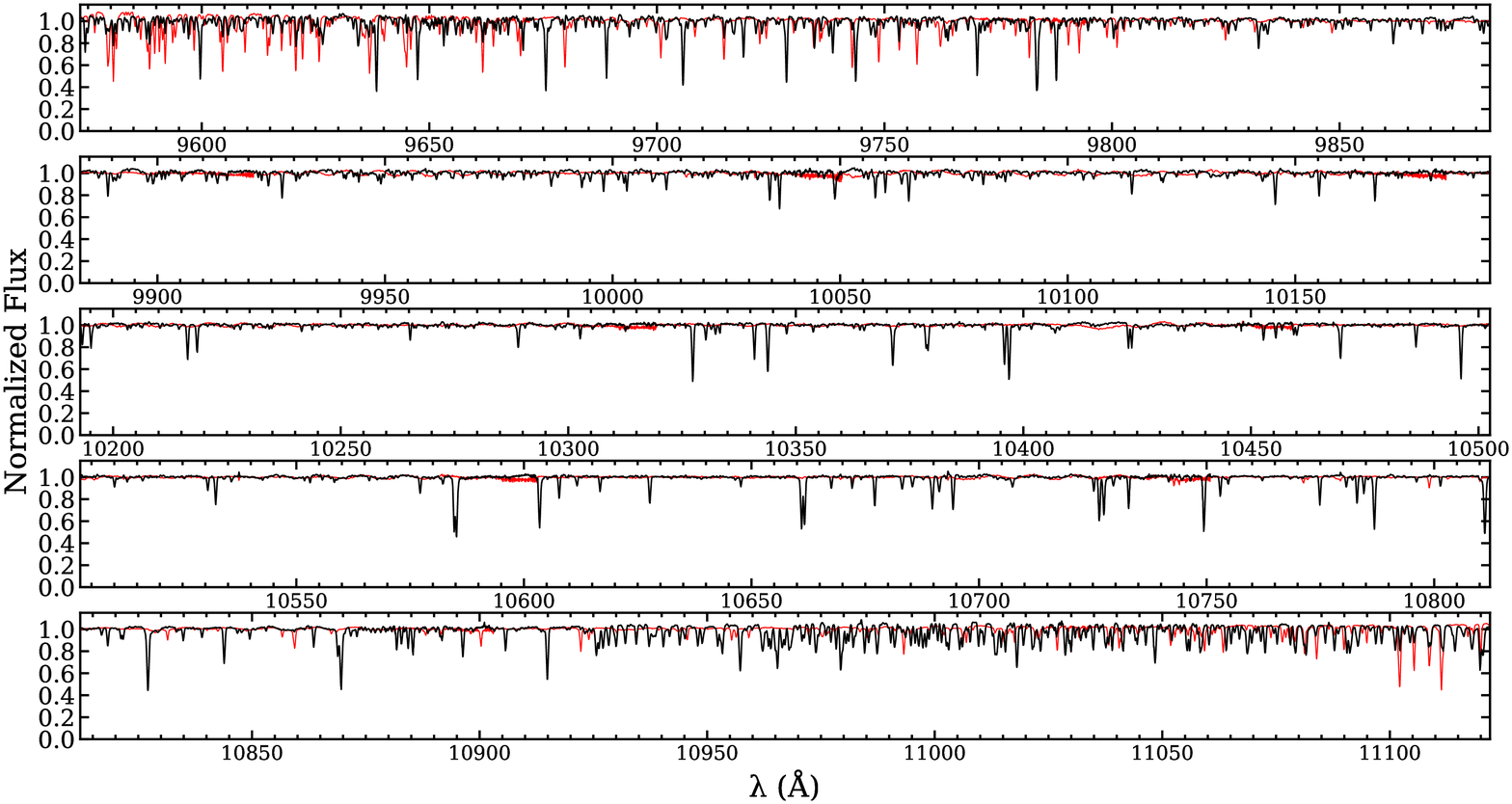}
    \caption{Observed Y-band Arcturus spectrum (black line) with the telluric correction (red line)}
    \label{yband}
\end{figure}
\end{landscape}
\begin{landscape}
\begin{figure}
    \includegraphics[width=25cm,height=15.5cm]{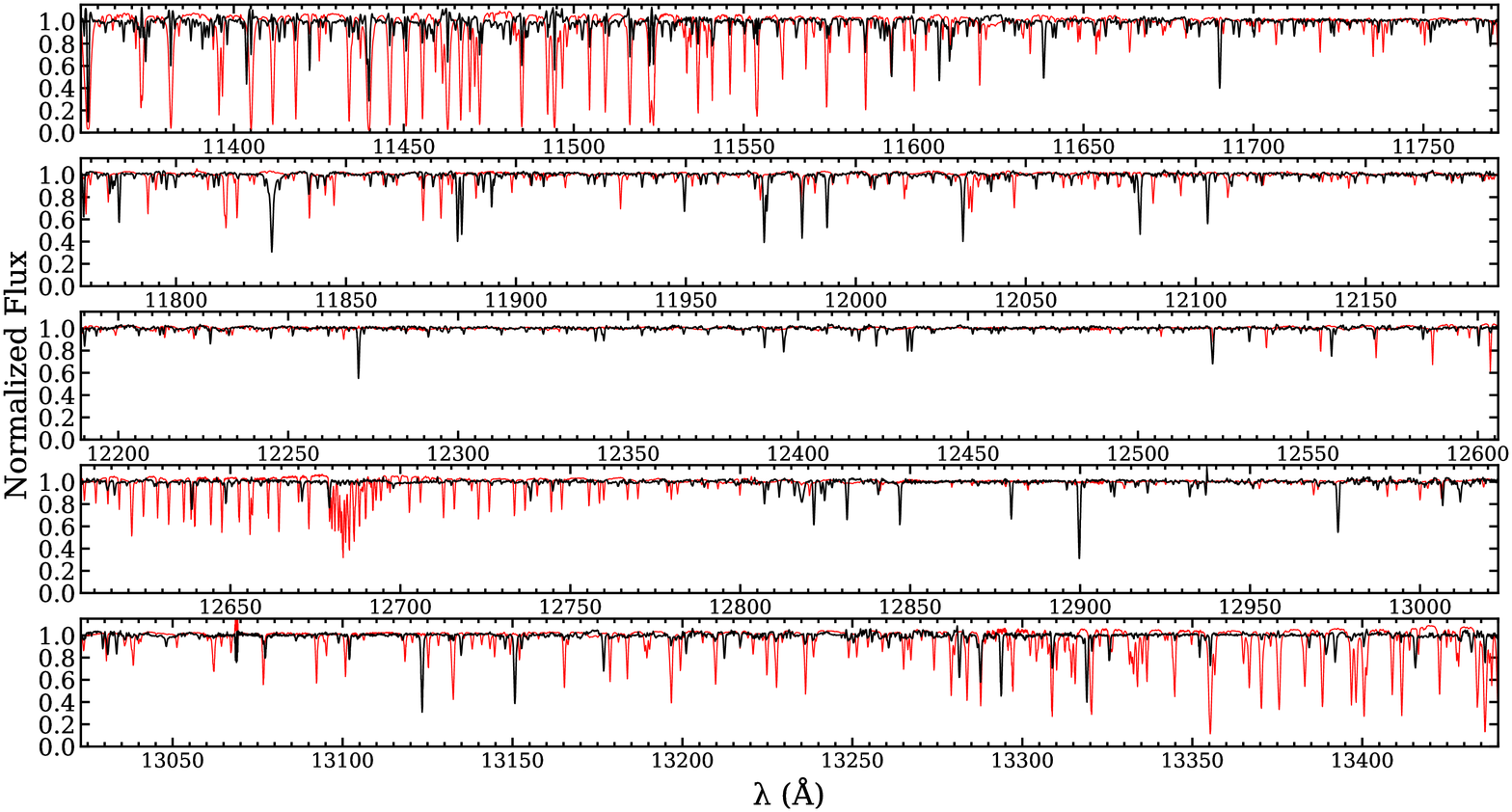}
    \caption{Observed J-band Arcturus spectrum (black line) with the telluric correction (red line)}
    \label{jband}
\end{figure}
\end{landscape}
\begin{landscape}
\begin{figure}
    \includegraphics[width=25cm,height=15.5cm]{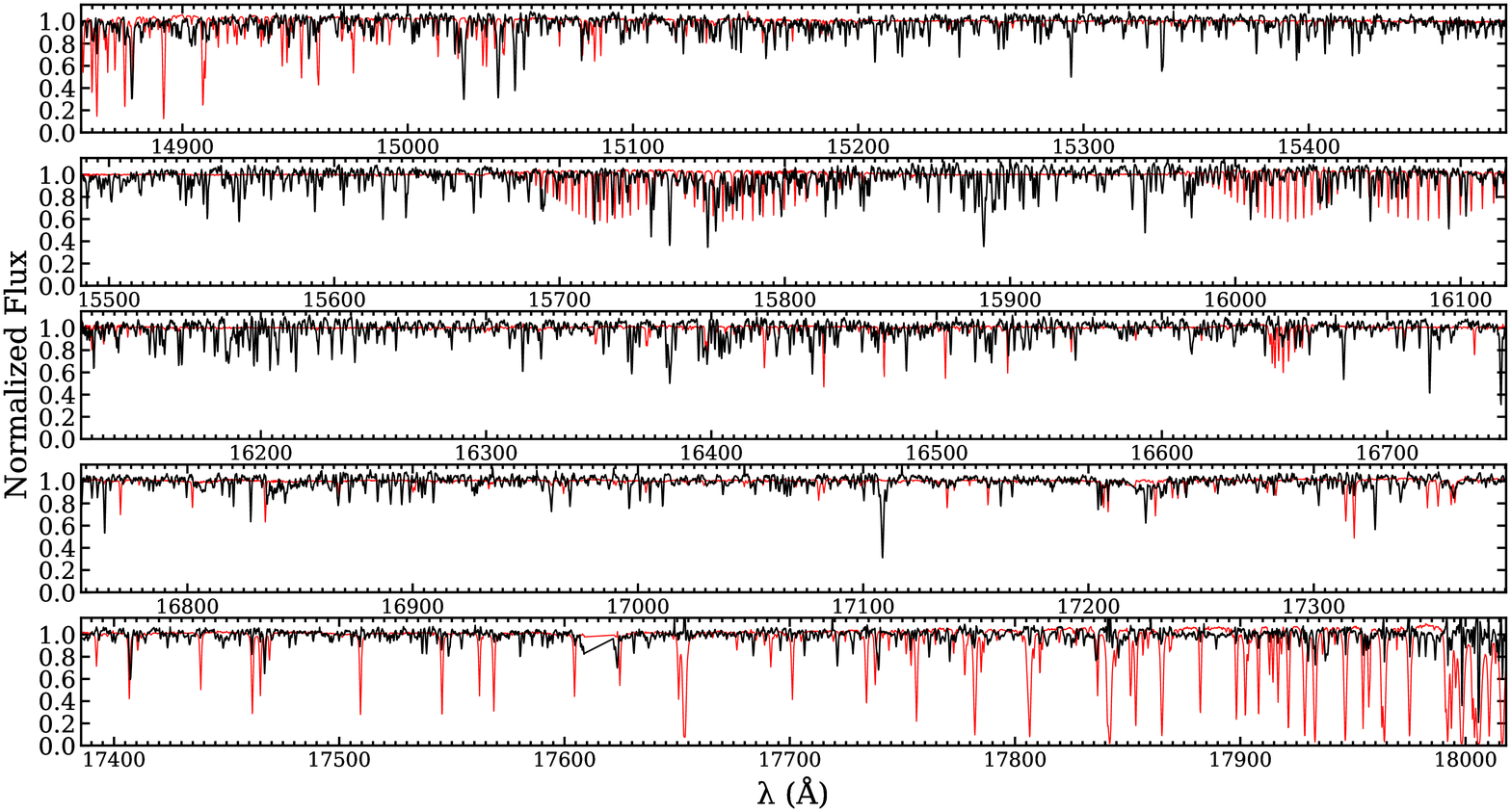}
    \caption{Observed H-band Arcturus spectrum (black line) with the telluric correction (red line)}
    \label{hband}
\end{figure}
\end{landscape}
\begin{landscape}
\begin{figure}
    \includegraphics[width=25cm,height=15.5cm]{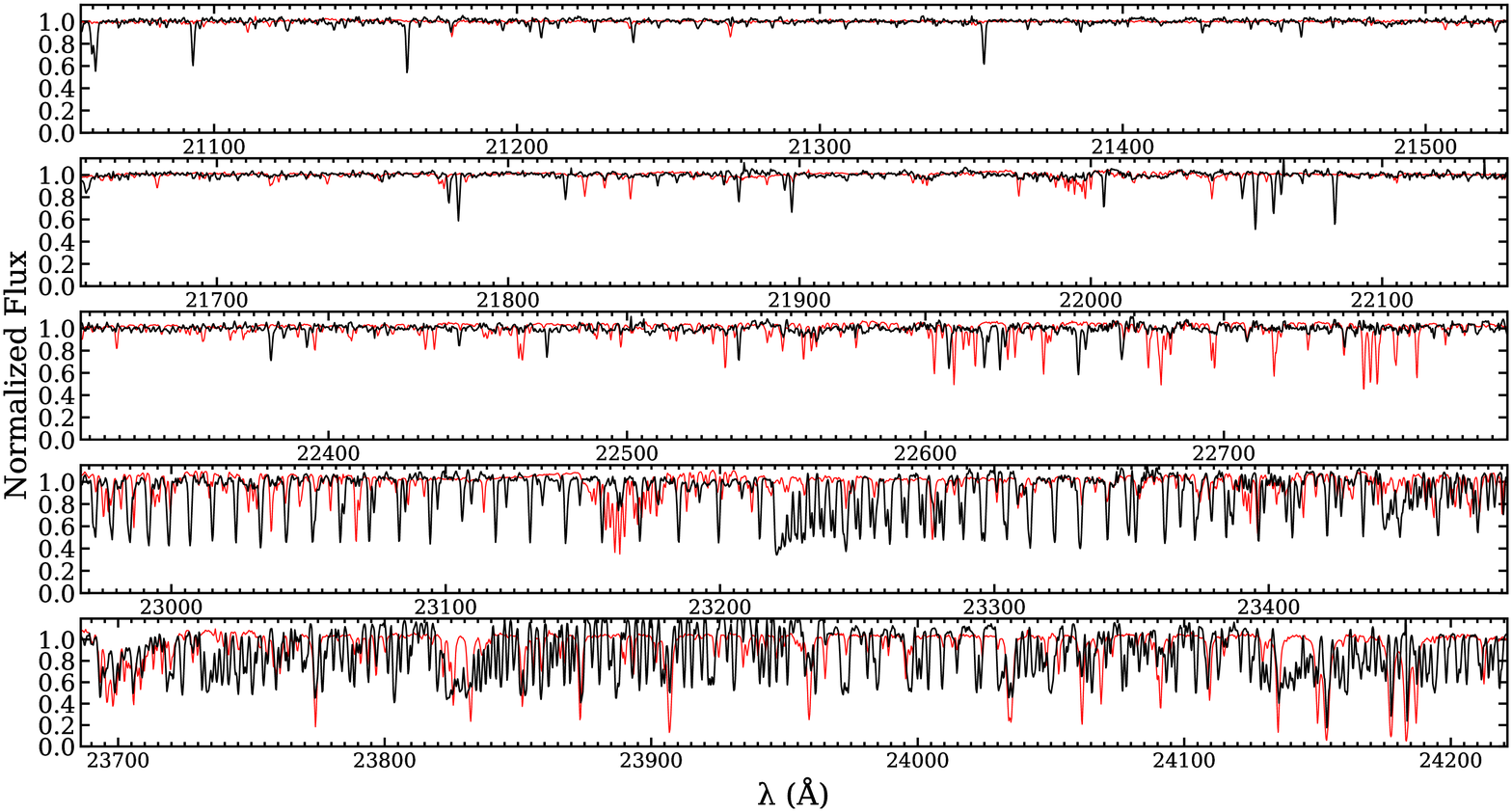}
    \caption{Observed K-band Arcturus spectrum (black line) with the telluric correction (red line)}
    \label{kband}
\end{figure}
\end{landscape}

\end{document}